\documentclass[11pt]{article}
\usepackage{graphicx}
\usepackage{latexsym,graphicx}
\usepackage{float}
\usepackage{amsmath}
\usepackage{amsfonts}
\usepackage{amssymb}
\usepackage{epstopdf}
\DeclareGraphicsExtensions{.eps}
\usepackage[left=2.5cm,top=2.5cm,right=2.5cm,bottom=2.5cm]{geometry}

\catcode `\@=11
\catcode `\@=12
\begin{document}
	\begin{center}
		\large{\bf{Modeling Transit Dark Energy in $f(R, L_m)$-gravity}} \\
		\vspace{5mm}
		\normalsize{Anirudh Pradhan$^1$, Dinesh Chandra Maurya$^{2}$, Gopikant K. Goswami$^3$, Aroonkumar  Beesham$^{4,5,6}$}\\
			\vspace{5mm}
		\normalsize{$^{1}$Centre for Cosmology, Astrophysics and Space Science (CCASS), GLA University, Mathura-281 406, Uttar Pradesh, India}\\
		\vspace{5mm}
		\normalsize{$^2$Center for Theoretical Physics and Mathematics, IASE (Deemed to be University), Sardarshahar-331 403 (Churu), Rajsthan, India. }\\
			\vspace{5mm}
		\normalsize{$^{3}$Department of Mathematics, Netaji Subhas University of Technology, Delhi, India}\\
			\vspace{5mm}
			\normalsize{$^{4}$Department of Mathematical Sciences, University of Zululand Private Bag X1001 Kwa-Dlangezwa 3886 South Africa}\\
			\normalsize{$^{5}$National Institute for Theoretical and Computational Sciences (NITheCS), South Africa}\\
			\normalsize{$^{6}$Faculty of Natural Sciences, Mangosuthu University of Technology, P O Box 12363, Jacobs 4052, South Africa}\\
			\vspace{2mm}
			$^1$E-mail: pradhan.anirudh@gmail.com \\
			\vspace{2mm}
			$^2$E-mail:dcmaurya563@gmail.com\\
			\vspace{2mm}
			$^3$E-mail: gk.goswami9@gmail.com \\
			\vspace{2mm}
			$^{4,5,6}$E-mail: abeesham@yahoo.com \\
	\end{center}

		\begin{abstract}

		This research paper deals with a transit dark energy cosmological model in $f(R, L_{m})$-gravity with observational constraints. For this, we consider a flat FLRW space-time and have taken a cosmological cosntant-like parameter $\beta$ in our field equations. The model has two energy parameters~ $\Omega_{m0}~ and~ \Omega_{\beta0}$, which govern the mechanism of the universe, in particular its present accelerated phase. To make the model cope with the present observational scenario, we consider three types of observational data set:  $46$ Hubble parameter data set,  SNe Ia $715$ data sets of distance modulus and apparent magnitude, and $40$ datasets of SNe Ia Bined compilation in the redshift $0\leq z<1.7$. We have approximated the present values of the energy parameters by applying $R^{2}$ and $\chi^{2}$-test in the observational and theoretical values of Hubble, distance modulus, and apparent magnitude parameters.  Also, we have measured the approximate present values of cosmographic  coefficients $\{H_{0}, q_{0}, j_{0}, s_{0}, l_{0}, m_{0}\}$. It is found that our approximated value-based model fits best with the observational module. We have found that as $t\to\infty$ (or $z\to 0$) then $\{q, j, s, l, m\}\to\{-1, 1, 1, 1, 1\}$. The cosmic age of the present universe is also approximated and comes up to the expectation. Our model shows a transit phase of the present accelerating universe with a deceleration in the past and has a transition point.
	\end{abstract}
	
	\smallskip
	\vspace{5mm}
	\vspace{1cm}


	\section{Introduction}
Cosmic observational studies like type Ia supernovae \cite{ref1, ref2,ref3}, Baryonic Acoustic Oscillations \cite{ref4,ref5}, Wilkinson Microwave Anisotropy Probe \cite{ref6}, Large scale Structure \cite{ref7,ref8} and the Cosmic Microwave Background Radiation \cite{ref9,ref10}  suggest that the expanding universe is in accelerating expansion phase. After that, among the theoretical cosmologists,  a thirst originated from developing cosmological models with accelerating phase expansion by either modifying Einstein's field equations or providing an alternative theory of gravity. The responsible components of the universe behind this acceleration have been known as ``Dark Energy" (DE) among researchers due to its mystery.
In the literature, several dark energy cosmological models were proposed to explain this acceleration. Conventional cosmology also strongly supports these ideas. The cosmological constant $\Lambda$-term, which is also researched as vacuum quantum energy, is considered to be the most likely candidate for dark energy \cite{ref11}. Even while the $\Lambda$-term closely matches the observed data, it still has two significant flaws, the fine-tuning problem and the cosmological constant problem (due to its origin)\cite{ref12}. The $ \Lambda $ term acquired from particle physics has a value over 120 orders of magnitude greater than that needed to match space observations. Imagining that Einstein's general theory of relativity model collapses on a vast cosmic scale, with more general factors defining the gravitational field, is another intriguing approach to explaining recent discoveries about the cosmic expansion scenario. The Einstein-Hilbert action of general relativity can be generalized in various ways.\\

The model introduced in \cite{ref13, ref14, ref15} where the default action has been changed the generic function $ f (R)$, where $ R $ is a Ricci scalar. $ f (R) $ gravity may be used to describe the late expansion scenario \cite{ref16}, and \cite{ref17, ref18} studied the limitations of a viable space model. There exists a good $ f (R) $ gravity models for solar system testing in \cite{ref19}-\cite{ref22}. \cite{ref23}-\cite{ref27} describe the observational aspects of the $ f (R) $ DE model, together with the solar system constraints and the equivalence principle of $ f (R) $ gravity. In \cite{ref28, ref29, ref30}, another $ f (R) $ model that combines early inflation with dark energy and undergoes local testing is detailed. In addition, one can consult the references \cite{ref31, ref32, ref33, ref33a, ref33b, ref33c, ref33d}  to learn more about the
$ f (R) $ gravity model's cosmic implications. \\

In \cite{ref34}, an extension of the $f(R)$ theory of gravity was presented, incorporating an explicit conjunction of the matter Lagrangian density $L_{m}$ and the generic function $f(R)$. During the non-geodesic motion of the heavy particles, extra forces perpendicular to the four-velocity vectors are created due to this matter-geometric connection. This model has been expanded to the situation of arbitrary matter-geometry pairings. In \cite{ref36}-\cite{ref40}, the cosmological and astrophysical ramifications of non-minimal matter-geometric couplings are examined in depth. Recently, Harko and Lobo \cite{ref41} introduced the $f(R,L_{m})$ gravity theory, where $f(R, L_{m})$ is an arbitrary function of the Lagrangian matter density $L_{m}$ and the Ricci scalar $R$. The $f(R, L_{m})$ gravitational theory is the most expansive of all gravitational theories developed in the Riemann space. In $f(R, L_{m})$ gravity theory, the motion of the test particle is non-geodesic, with extra forces orthogonal to the four-velocity vectors. The gravitational model $f(R,L m)$ permits explicit breaches of the equivalence principle, which are tightly confined by the solar system test \cite{ref42,ref43}. Wang and Liao \cite{ref44} have recently examined the energy condition of $f(R, L_{m})$ gravity. Gonclaves and Moraes employ the $f(R, L_{m})$ gravity model \cite{ref45}. We explored cosmology by combining non-minimum matter geometries with a specific version of the $f(R, L_{m})$ function  \cite{ref46}. \\

This study examines an observationally constrained transit dark energy cosmology model in $f(R, L_{m})$-gravity. We consider a flat FLRW space-time for this and have included a term in our field equations called $\beta$ that resembles a cosmological constant. $\Omega_{m0}~ and~ \Omega_{\beta0}$, two energy parameters in the model, determine the universe's mechanism, particularly its current accelerated phase. We take into account three different types of observational data sets to help the model adapt to the current state of observable phenomena: $46$ Hubble parameter data set, SNe Ia $715$ data sets of distance modulus and apparent magnitude and $40$ datasets of SNe Ia Bined compilation in the redshift $0\leq z<1.7$. Using $R^{2}$ and $\chi^{2}$-test, the observational and theoretical values of the Hubble, distance modulus, and apparent magnitude parameters, we have approximated the current values of the energy parameters.
We have also calculated the estimated present values of the cosmographic coefficients $\{H_{0}, q_{0}, j_{0}, s_{0}, l_{0}, m_{0}\}$. Our approximation-based model was found to fit the observational module the best. As $t\to\infty$ (or $z\to 0$) , we have discovered that $\{q, j, s, l, m\}\to\{-1, 1, 1, 1, 1\}$. In our scenario, there is a transition point and a transit phase of an accelerating cosmos that decelerated in the past. The current universe's cosmic age is likewise approximated and meets expectations. Our model has a transition point and depicts a transit phase of the now accelerating world with a previously decelerating universe. \\

This investigation is divided into the following sections: In section $2$, we have mentioned the formulation of $f(R, L_{m})$ gravity. In section 3, field equations are obtained for a flat FLRW space-time universe using a perfect-fluid stress-energy momentum tensor. In section $4$, we have obtained the cosmological solutions for $f(R, L_{m})=\frac{R}{2}+L_{m}^{n}-\beta$ model. In section $5$, we made an observational constraint on Hubble parameter and distance modulus function with observational $H_{0}$ data sets, SNe Ia $715$ data sets, and SNe Ia $40$ Bined data sets. Section $6$ contains result analysis with cosmographic coefficients and the age of the present universe. Finally, in the last section, $7$ conclusions are included.

\section{Modified $f(R, L_{m})$-gravity theory}
The action principle for $f(R, L_{m})$-gravity is taken as
\begin{equation}\label{eq1}
  I=\int{f(R, L_{m})}\sqrt{-g}d^{4}x
\end{equation}
where $L_{m}$ is the matter Lagrangian of the perfect fluid, $R$ is the Ricci scalar curvature and these are related by an arbitrary function $f(R, L_{m})$. The Ricci-scalar $R$ is defined in terms of metric tensor $g^{ij}$ and Ricci-tensor $R_{ij}$ as given below
\begin{equation}\label{eq2}
  R=g^{ij}R_{ij}
\end{equation}
where the Ricci-tensor is given by
\begin{equation}\label{eq3}
  R_{ij}=-\frac{\partial^{2}}{\partial x^{i}\partial x^{j}}\log\sqrt{|g|}+\partial_{k}\Gamma^{k}_{ij}-\Gamma^{\alpha}_{ik}\Gamma^{k}_{\alpha j}+\Gamma^{\alpha}_{ij}\Gamma^{k}_{\alpha k}
\end{equation}
in terms of well-known Levi-Civita connection $\Gamma^{i}_{jk}$ which is defined as
\begin{equation}\label{eq4}
  \Gamma^{\alpha}_{\beta \gamma}=\frac{1}{2}g^{\alpha k}\left(\frac{\partial{g_{\gamma k}}}{\partial{x^{\beta}}}+\frac{\partial{g_{k\beta}}}{\partial{x^{\gamma}}}-\frac{\partial{g_{\beta\gamma}}}{\partial{x^{k}}} \right)
\end{equation}
On the variation of action (\ref{eq1}) over metric tensor $g_{ij}$, one can find the following equations
\begin{equation}\label{eq5}
  f_{R}R_{ij}+(g_{ij}\nabla_{k}\nabla^{k}-\nabla_{i}\nabla_{j})f_{R}-\frac{1}{2}(f-f_{L_{m}}L_{m})g_{ij}=\frac{1}{2}f_{L_{m}}T_{ij}
\end{equation}
where $f_{R}=\frac{\partial{f}}{\partial{R}},~~f_{L_{m}}=\frac{\partial{f}}{\partial{L_{m}}}$ and $T_{ij}$ is the Stress-energy momentum tensor for perfect-fluid, defined by
\begin{equation}\label{eq6}
  T_{ij}=\frac{-2}{\sqrt{-g}}\frac{\delta{(\sqrt{-g}L_{m})}}{\delta{g^{ij}}}
\end{equation}
Now, contracting the field equation (\ref{eq5}), we get the relation between Ricci-scalar curvature $R$, matter Lagrangian density $L_{m}$ and $T$ the trace of the stress-energy-momentum tensor $T_{ij}$ as
\begin{equation}\label{eq7}
  Rf_{R}+3\Box f_{R}-2(f-f_{L_{m}}L^{m})=\frac{1}{2}f_{L_{m}}T
\end{equation}
where $\Box{F}=\frac{1}{\sqrt{-g}}\partial_{\alpha}(\sqrt{-g}g^{\alpha\beta}\partial_{\beta}F)$ for any function $F$.\\
Taking covariant derivative of Eq.~(\ref{eq5}), we obtain
\begin{equation}\label{eq8}
  \nabla^{i}T_{ij}=2\nabla^{i}log(f_{L_{m}})\frac{\partial{L_{m}}}{\partial{g^{ij}}}
\end{equation}

\section{Field Equations}
For the spatial isotropic and homogeneous universe, we consider the following flat Friedman-Lamatre-Robertson-Walker (FLRW) metric \cite{ref47} to analyzed the present cosmological model,
\begin{equation}\label{eq9}
  ds^{2}=-dt^{2}+a^{2}(t)[dx^{2}+dy^{2}+dz^{2}]
\end{equation}
Where, $a(t)$ is known as the scale factor which represents the characteristics of the expanding universe at any instant. For the metric (\ref{eq9}), the non-zero components of Christoffel symbols are
\begin{equation}\label{eq10}
  \Gamma^{0}_{ij}=-\frac{1}{2}g^{00}\frac{\partial g_{ij}}{\partial x^{0}},~~\Gamma^{k}_{0j}=\Gamma^{k}_{j0}=\frac{1}{2}g^{k\lambda}\frac{\partial g_{j\lambda}}{\partial x^{0}}
\end{equation}
where, $i, j, k=1, 2, 3$.\\

Equation (\ref{eq3}) allows us to determine the Ricci curvature tensor's non-zero components as
\begin{equation}\label{eq11}
  R_{0}^{0}=3\frac{\ddot{a}}{a},~~R_{1}^{1}=R_{2}^{2}=R_{3}^{3}=\frac{\ddot{a}}{a}+2\left(\frac{\dot{a}}{a}\right)^{2}
\end{equation}
As a result, the line element's associated Ricci-scalar ($R$) is found to be
\begin{equation}\label{eq12}
  R=6\frac{\ddot{a}}{a}+6\left(\frac{\dot{a}}{a}\right)^{2}=6(\dot{H}+2H^{2})
\end{equation}
where $H$ is the Hubble parameter defined by $H=\frac{\dot{a}}{a}$.\\
The stress-energy momentum tensor for the perfect-fluid matter filled in the universe corresponding to the line element (\ref{eq9}) is taken as
\begin{equation}\label{eq13}
  T_{ij}=(\rho+p)u_{i}u_{j}+pg_{ij}
\end{equation}
or
\begin{equation}\nonumber
  T_{i}^{i}=diag[-\rho, p, p, p]
\end{equation}
where $p$ is the pressure of the cosmic fluid, $\rho$ is the energy density, $g^{ij}$ is the metric tensor and $u^{i}=(1,0,0,0)$ components of co-moving four velocity vectors in cosmic fluid such that $u_{i}u^{i}=-1$.\\
The modified Friedmann equations that describe the universe's dynamics in $f(R, L_{m})$ gravity are as follows:
\begin{equation}\label{eq14}
  R_{0}^{0}f_{R}-\frac{1}{2}(f-f_{L_{m}}L_{m})+3H\dot{f_{R}}=\frac{1}{2}f_{L_{m}}T_{0}^{0}
\end{equation}
and
\begin{equation}\label{eq15}
  R_{1}^{1}f_{R}-\frac{1}{2}(f-f_{L_{m}}L_{m})-\ddot{f_{R}}-3H\dot{f_{R}}=\frac{1}{2}f_{L_{m}}T_{1}^{1}
\end{equation}

\section{Cosmological Solutions for $f(R, L_{m})$-Gravity}
In this study, we have considered the $f(R, L_{m})$-gravity \cite{ref47} of the form
\begin{equation}\label{eq16}
  f(R, L_{m})=\frac{R}{2}+ \alpha L_{m}^{n}-\beta
\end{equation}
where $n$,$\alpha$ and $\beta$ are arbitrary constants. $\alpha$ is coupling constant. Its dimension is that of $L_m^{n-1}$ \\
Thus, for this particular form of $f(R, L_{m})$-gravity model, we have considered dusty universe with $L_{m}=\rho$ \cite{ref49} and thus, for the matter-dominated universe, the Friedmann equations (\ref{eq14}) and (\ref{eq15}) become
\begin{equation}\label{eq17}
  3H^{2}=(2n-1)\alpha\rho^{n}+\beta
\end{equation}
and
\begin{equation}\label{eq18}
  2\dot{H}+3H^{2}=(n-1)\alpha\rho^{n}+\beta
\end{equation}

After taking a trace of the field equations, one can also get the following matter conservation equation
\begin{equation}\label{eq19}
  (2n-1)\dot{\rho}+3H\rho=0
\end{equation}
In particular, one can obtain the standard Friedmann equations of GR for $n=1$,$\alpha=1$ and $\beta=0$ \\

Solving Eq.~(\ref{eq19}), we get energy density of matter fluid as
\begin{equation}\label{eq20}
  \rho=\rho_{0}\left(\frac{a_{0}}{a}\right)^{\frac{3}{2n-1}}
\end{equation}
where $a_{0}$ and $\rho_{0}$ are present values of scale factor and energy density respectively.\\
From equation (\ref{eq17}), we can find
\begin{equation}\label{eq21}
  \Omega_{m}+\Omega_{\beta}=\frac{1}{2n-1}
\end{equation}
where $\Omega_{m}=\frac{\alpha\rho^{n}}{3H^{2}}$ and $\Omega_{\beta}=\frac{\beta}{3(2n-1)H^{2}}$ are called as matter energy density parameter and dark energy density parameter respectively.\\
Using Eq.~(\ref{eq20}) in (\ref{eq21}), we get the Hubble parameter as
\begin{equation}\label{eq22}
  H=H_{0}\sqrt{2n-1}\sqrt{\Omega_{m0}\left(\frac{a_{0}}{a}\right)^{\frac{3n}{2n-1}}+\Omega_{\beta0}}
\end{equation}
and in terms of redshift $z$, using $\frac{a_{0}}{a}=1+z$ as given in \cite{ref50} in (\ref{eq22}), we get
\begin{equation}\label{eq23}
  H(z)=H_{0}\sqrt{2n-1}\sqrt{\Omega_{m0}(1+z)^{\frac{3n}{2n-1}}+\Omega_{\beta0}}
\end{equation}
with $\Omega_{m0}$, $\Omega_{\beta0}$ and $H_{0}$ are the present values of corresponding parameters $\Omega_{m}$, $\Omega_{\beta}$ and $H$ respectively i.e  $\Omega_{m0}=\frac{\alpha\rho_0^{n}}{3H_0^{2}}$ and $\Omega_{\beta 0}=\frac{\beta}{3(2n-1)H_0^{2}}.$

Eq. (\ref{eq22}) is integrated, and the scale-factor is obtained as
\begin{equation}\label{eq24}
  a(t)=k_{0}[sinh(k_{1}t+k_{2})]^{\frac{2(2n-1)}{3n}}
\end{equation}
where $k_{0}=a_{0}\left(\frac{\Omega_{m0}}{\Omega_{\beta0}}\right)^{\frac{2n-1}{3n}}$, $k_{1}=\frac{2}{3n}H_{0}\sqrt{\Omega_{\beta0}(2n-1)^{3}}$ and $k_{2}$ is an integrating constant.

We can solve Eq.~(\ref{eq21}), (\ref{eq22}) and (\ref{eq18}) to get deceleration parameter $q$ as

\begin{equation}\label{eq26}
  q(z)=-1+\frac{3n}{2(2n-1)}-\frac{3n}{2(2n-1)}\frac{\Omega_{\beta0}}{\Omega_{m0}(1+z)^{\frac{3n}{2n-1}}+\Omega_{\beta0}}
\end{equation}

The transition point $z_{t}$ for which $q(z_{t})=0$ obtained as
\begin{equation}\label{eq26a}
	z_{t}=\left[\frac{2n-2}{2-n}\frac{\Omega_{\beta0}}{\Omega_{m0}}\right]^{\frac{2n-1}{3n}}-1
\end{equation}

For  $z>z_{t}$, universe was in a decelerating phase, and for $z<z_{t}$, it is accelerating. Thus, as suggested in recent observations \cite{ref1}-\cite{ref10}, our resulting model depicts an expanding universe in the transit phase (decelerating to accelerating phase).


\section{Observational Constraints}
To validate and viability of our model, we have compared our results with observational data sets through the Hubble parameter $H(z)$ with redshift. Now, using Eq.~(\ref{eq24}) with the relation $\frac{a_{0}}{a}=1+z$ given in \cite{ref50}, for our model, we have determined the following relationship between cosmic time $t$ and redshift $z$:

\begin{equation}\label{eq27}
  t(z)=\frac{1}{k_{1}}sinh^{-1}\left[\left(\frac{a_{0}/k_{0}}{1+z}\right)^{\frac{3n}{2(2n-1)}}\right]-\frac{k_{2}}{k_{1}}
\end{equation}
The formulation for the cosmic time $t(z)$ in terms of redshift $z$ is represented by equation (\ref{eq27}). One might think of this as the current age of the universe, which is $z\to\infty$  then $t\to0$ and as $z\to0$ then $t\to t_{0}$.

\subsection{Hubble Parameter}
At present days cosmological studies expect researchers in this field to validate their theoretical results with observational data sets obtained from various observatories and study centers, viz. \cite{ref1}-\cite{ref10}. Therefore, to validate our derived model, we have found the best fit values of model parameters by comparing our model with observational datasets, we have considered $46$ Hubble data sets $H(z)$ with high redshift $(0 \leq z \leq 2.36)$ estimated by different cosmologists \cite{ref51}-\cite{ref57} \& \cite{ref61}-\cite{ref70} as mentioned in Table $2$. With the aid of the $R^{2}$ formula, we have employed a strategy in the current study to estimate the present values of $\Omega_{m0}$, $\Omega_{\beta0}$, $H_{0}$, $n$ by comparing the theoretical and observed results.

\begin{equation}\nonumber
	R^{2}_{H}=1-\frac{\sum_{i=1}^{46}[(H_{i})_{ob}-(H_{i})_{th}]^{2}}{\sum_{i=1}^{46}[(H_{i})_{ob}-(H_{i})_{mean}]^{2}}
\end{equation}

Here, $(H_{i})_{th}$ is the theoretical value of the Hubble parameter $H(z)$ as given in Eq.~(\ref{eq23}), $(H_{i})_{ob}$ is the observational values of $H(z)$ (as mentioned in Table $2$) and $(H_{i})_{mean}$ is the mean of Hubble data $(H_{i})_{ob}$ (mentioned in Table $2$).\\

To validate our model in front of observational data, we have used the $R^{2}$-test formula nearest to the ideal case $R^{2}=1$ the condition in which both the data sets (observational and theoretical) are compatible. Hence, we have estimated the best fit values of $\Omega_{m0}, \Omega_{\beta0}, H_{0}, n$ as mentioned below in Table $1$ with $R^{2}=0.945364$ at $95\%$ confidence level of bounds. Thus, we have estimated the present value of the Hubble parameter $H_{0}=68.9596$ km/s/Mpc. In 2018, the Planck Collaboration determined the Hubble constant to be $H_{0}=67.4\pm0.5$ km/s/Mpc using the Cosmic Microwave Background (CMB) temperature and polarisation anisotropies and the $\Lambda$CDM cosmological model. In contrast, the SH0ES Collaboration (Riess et al.\cite{ref59}) determined the Hubble constant to be $H_{0}=73.2\pm1.3$ km/s/Mpc using Cepheids and supernovae. Cunha et al. \cite{ref60} estimated the value of Hubble constant $H_{0}\approx74$ km/s/Mpc. The best fit curve of Hubble parameter $H(z)$ is represented in Figure 1. Figure 2 represents $1 \sigma$ and $2 \sigma$ confidence regions for estimated parameters $\Omega_{m0}$  and Hubble constant $H_0$ for our model.

\begin{table}[H]
	\centering
	\begin{tabular}{|c|c|}
		\hline
		Parameter          & Values    \\
		\hline
		\\
		$\Omega_{m0}$      & 0.306973$^{+0.0174}_{-0.0107}$  \\
		\\
		
		$\Omega_{\beta0}$  & 0.618952$^{+ 0.908526}_{-0.915226}$  \\
		\\
		
		$H_{0}$            & 68.9596$^{+2.17}_{-0.9538}$ \\ 
		
		\\
		
		$n$                & 1.04    \\
		\\
		$R^{2}$            & 0.945364   \\
		
		\hline
	\end{tabular}
	\caption{For $46$ Hubble parameter $H(z)$ data sets, the best fit values of $\Omega_{m0}, \Omega_{\beta0}, H_{0}, n$ using the $R^{2}$-test at the $95\%$ confidence level of bounds.}\label{T1}
\end{table}


\begin{figure}[ht]
	\begin{center}
	a. \includegraphics[scale=0.40]{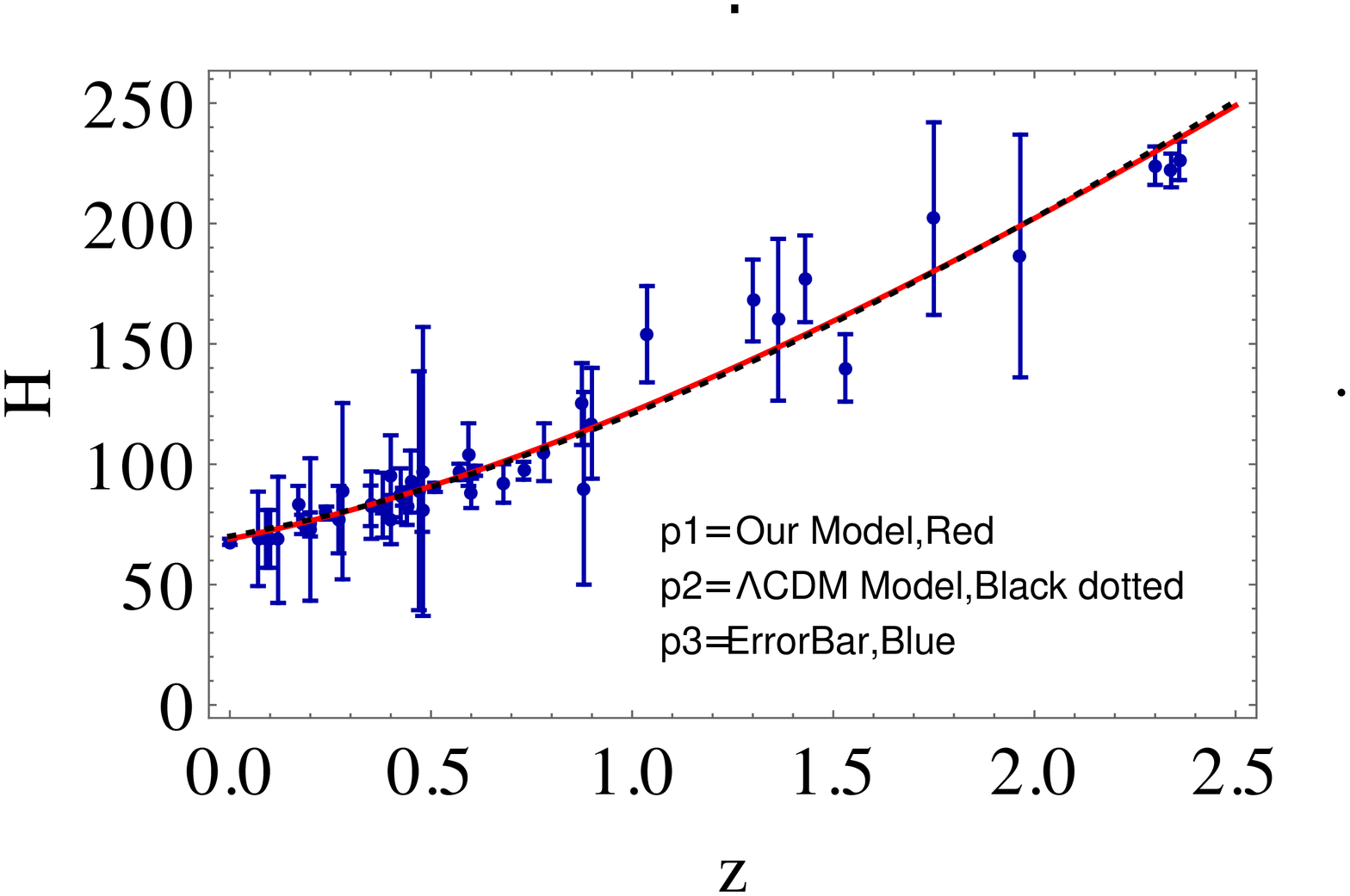}
	b. \includegraphics[scale=0.70]{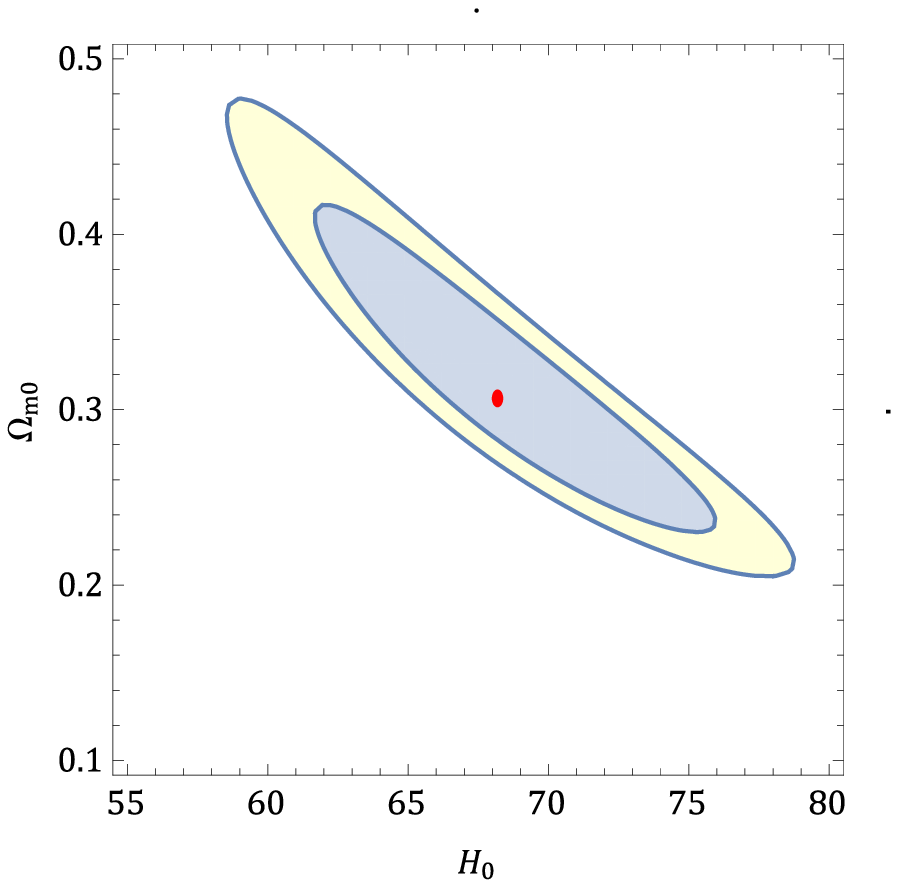}
	\end{center}
	\caption{(a) Hubble parameter $H(z)$ Plot for the best fit values of $\Omega_{m0}, \Omega_{\beta0}, H_{0}, n$  with $H_{0}$ data sets, using $R^{2}$-test at $95\%$ confidence level of bounds as mentioned in Table $1$. (b) $1 \sigma$ and $2 \sigma$ confidence regions for estimated parameters $\Omega_{m0}$  and Hubble constant $H_0$  }
\end{figure}

\begin{figure}[ht]
	\begin{center}
	a. \includegraphics[scale=0.90]{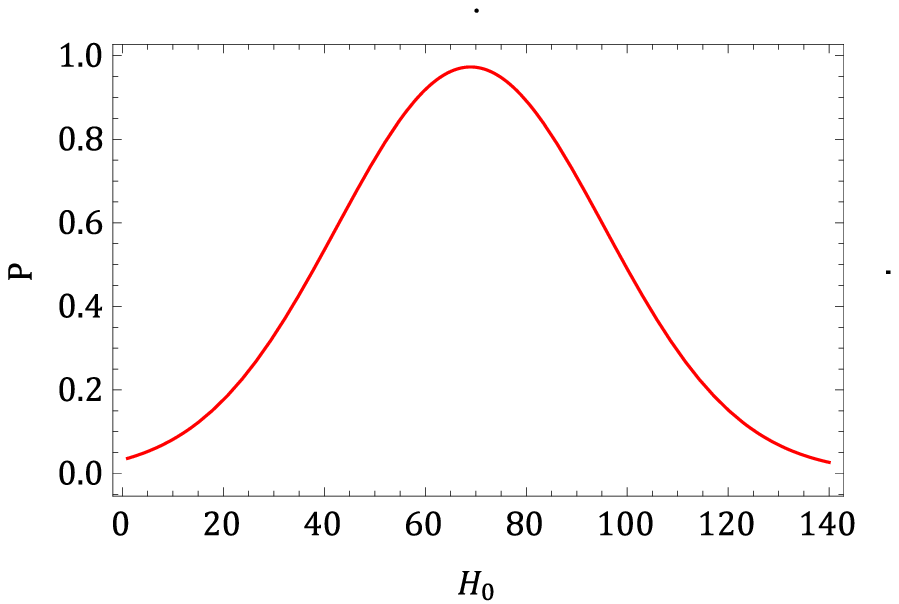}
	b. \includegraphics[scale=0.50]{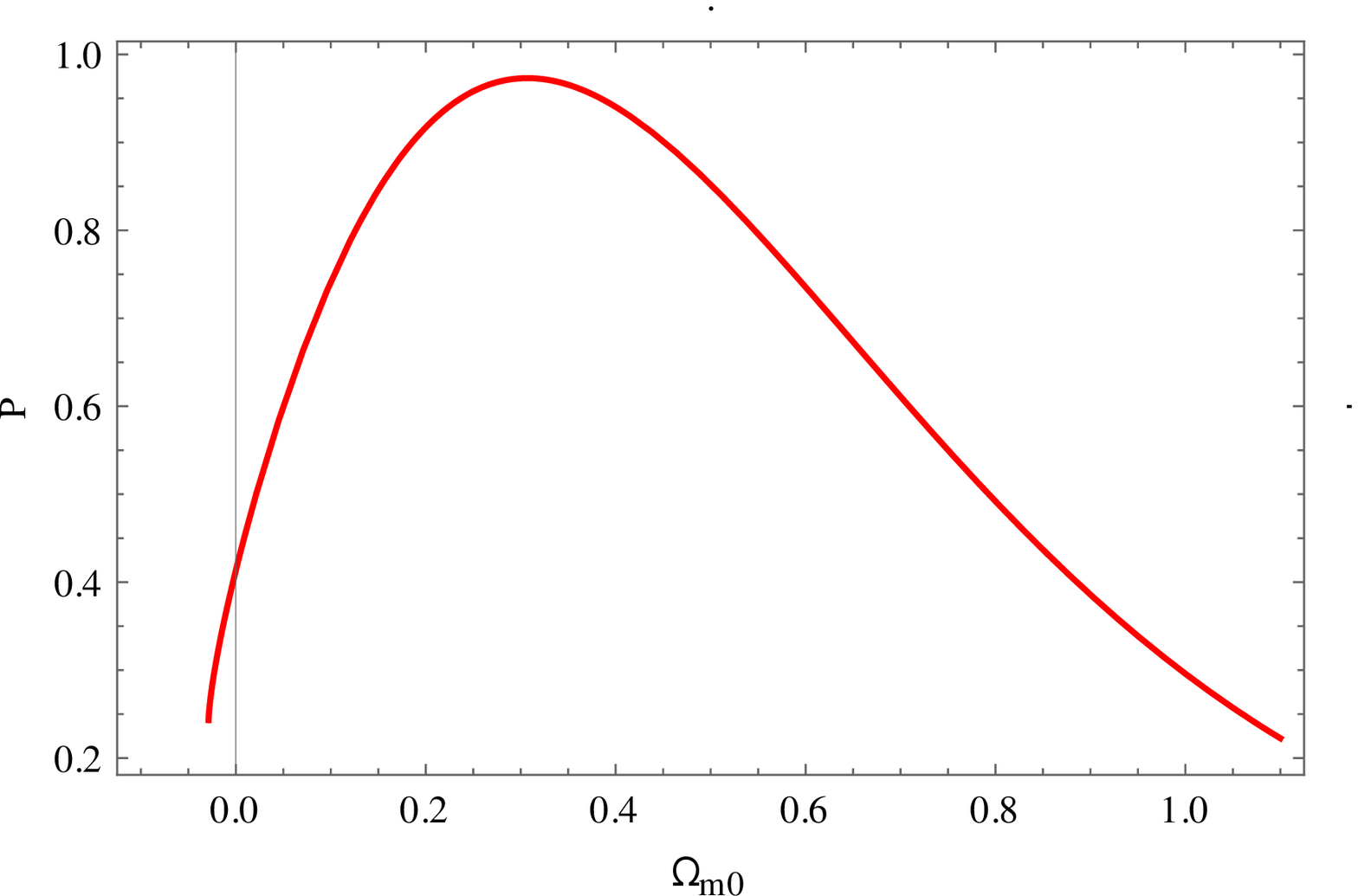}
	\end{center}
	\caption{(a,b)Likelyhood plots for estimated Hubble $H_0$ and barion energy parameter $\Omega_{m0}$   }
\end{figure}
\newpage
\begin{table}[H]
	\centering
	\begin{tabular}{|c|c|c|c|c|c|c|c|c|c|}
		\hline
		S.No. & $z$   & $H(z)$  & $\sigma_{H}$  & Reference  & S.No. & $z$   & $H(z)$  & $\sigma_{H}$  & Reference\\
		\hline
		1  & $0$      & $67.77$ & $1.30$   & \cite{ref51}  & 24  & $0.4783$  & $80.9$  & $9$     & \cite{ref69}\\
		2  & $0.07$   & $69$    & $19.6$   & \cite{ref52}  & 25  & $0.48$    & $97$    & $60$    & \cite{ref53}\\
		3  & $0.09$   & $69$    & $12$     & \cite{ref68}  & 26  & $0.51$    & $90.4$  & $1.9$   & \cite{ref55}\\
		4  & $0.10$   & $69$    & $12$     & \cite{ref53}  & 27  & $0.57$    & $96.8$  & $3.4$   & \cite{ref70}\\
		5  & $0.12$   & $68.6$  & $26.2$   & \cite{ref52}  & 28  & $0.593$   & $104$   & $13$    & \cite{ref67}\\
		6  & $0.17$   & $83$    & $8$      & \cite{ref53}  & 29  & $0.60$    & $87.9$  & $6.1$   & \cite{ref61}\\
		7  & $0.179$  & $75$    & $4$      & \cite{ref67}  & 30  & $0.61$    & $97.3$  & $2.1$   & \cite{ref55}\\
		8  & $0.1993$ & $75$    & $5$      & \cite{ref67}  & 31  & $0.68$    & $92$    & $8$     & \cite{ref67}\\
		9  & $0.2$    & $72.9$  & $29.6$   & \cite{ref52}  & 32  & $0.73$    & $97.3$  & $7$     & \cite{ref61}\\
		10  & $0.24$   & $79.7$  & $2.7$    & \cite{ref54}  & 33  & $0.781$   & $105$   & $12$    & \cite{ref67}\\
		11  & $0.27$   & $77$    & $14$     & \cite{ref53}  & 34  & $0.875$   & $125$   & $17$    & \cite{ref67}\\
		12  & $0.28$   & $88.8$  & $36.6$   & \cite{ref52}  & 35  & $0.88$    & $90$    & $40$    & \cite{ref53}\\
		13  & $0.35$   & $82.7$  & $8.4$    & \cite{ref56}  & 36  & $0.9$     & $117$   & $23$    & \cite{ref53}\\
		14  & $0.352$  & $83$    & $14$     & \cite{ref67}  & 37  & $1.037$   & $154$   & $20$    & \cite{ref54}\\
		15  & $0.38$   & $81.5$  & $1.9$    & \cite{ref55}  & 38  & $1.3$     & $168$   & $17$    & \cite{ref53}\\
		16  & $0.3802$ & $83$    & $13.5$   & \cite{ref56}  & 39  & $1.363$   & $160$   & $33.6$  & \cite{ref63}\\
		17  & $0.4$    & $95$    & $17$     & \cite{ref68}  & 40  & $1.43$    & $177$   & $18$    & \cite{ref53}\\
		18  & $0.004$  & $77$    & $10.2$   & \cite{ref69}  & 41  & $1.53$    & $140$   & $14$    & \cite{ref53}\\
		19  & $0.4247$ & $87.1$  & $11.2$   & \cite{ref69}  & 42  & $1.75$    & $202$   & $40$    & \cite{ref63}\\
		20  & $0.43$   & $86.5$  & $3.7$    & \cite{ref54}  & 43  & $1.965$   & $186.5$ & $50.4$  & \cite{ref54}\\
		21  & $0.44$   & $82.6$  & $7.8$    & \cite{ref61}  & 44  & $2.3$     & $224$   & $8$     & \cite{ref66}\\
		22  & $0.44497$& $92.8$  & $12.9$   & \cite{ref69}  & 45  & $2.34$    & $222$   & $7$     & \cite{ref64}\\
		23  & $0.47$   & $89$    & $49.6$   & \cite{ref62}  & 46  & $2.36$    & $226$   & $8$     & \cite{ref65}\\
		\hline
	\end{tabular}
	\caption{Hubble's constant table.}\label{T2}
\end{table}
 \subsection{Luminosity Distance} Cosmology has emerged as a significant area of high-level research. The presence of dark matter and dark energy in abundance in the universe are explored under it. The phenomena of gravitating licensing have roots in dark matter.
 The redshift-luminosity distance relation, measurements of the distance modulus $\mu$, and apparent magnitude $m_{b}$ have emerged as powerful observational tools to understand the universe's development. After accounting for redshift, the luminosity distance ($D_L$) is the distance to a far-off luminous object. The luminosity distance is used to calculate a source's flux. It is provided as

 \begin{equation}\label{eq28}
	D_{L}=a_{0} r (1+z).
\end{equation}
where the source's radial coordinate is represented by $r$. Consider of a light ray's path as
\begin{equation}\label{eq29}
	\frac{d\theta}{ds}=0 ~ ~ \mbox{and} ~ ~ \frac{d\phi}{ds}=0,
\end{equation}
We get the followings from the Geodesic of light photons
\begin{equation}\label{eq30}
	\frac{d^2\theta}{ds^2}=0 ~ ~ \mbox{and} ~ ~ \frac{d^2\phi}{ds^2}=0.
\end{equation}
  Radial directional ray continues to move along the $r$-direction always, and we get the following  light ray path
\begin{equation}\label{eq31}
	ds^{2}= c^{2}dt^{2}- \frac{a^{2}}{1+kr^2}dr^2=0.
\end{equation}
 Taking $k=0$.  we obtain
\begin{equation}\label{eq32}
	r  =  \int^r_{0}dr = \int^t_{0}\frac{cdt}{a(t)} = \frac{1}{a_{0}H_{0}}\int^z_0\frac{cdz}{h(z)}
\end{equation}

where we have used $ dt=dz/\dot{z}, \dot{z}=-H(1+z)~\&~ h(z)=\frac{H}{H_0}.$

This gives  the luminosity distance  as:

\begin{equation}\label{33}
	D_{L}=\frac{c(1+z)}{H_{0}}\int^z_0\frac{dz}{h(z)}
\end{equation}

\subsection{ Apparent Magnitude $m_{b}$ and Distance Modulus $\mu$ }
The apparent magnitude $m_{b}$ and distance modulus $\mu$ are the two most required obsrvational parameters as the leading resources of observational cosmology. SN Ia union 2.1 compilation data set provide $\mu$ and  $m_{b}$ of 715 low and high red shift SN Ia where as  pantheon compilation data set provide  $m_{b}$ of 1120 low and high red shift SN Ia. These parameters have following theoretical formulisms

\begin{eqnarray*}
	\mu & = &  m_{b}-M     \\
	& = &  5log_{10}\left(\frac{D_L}{Mpc}\right)+25 \\
	& = & 25+  5log_{10}\left[\frac{c(1+z)}{H_0} \int^z_0\frac{dz}{h(z)}\right].
\end{eqnarray*}
\begin{equation}\label{eq34}
\end{equation}
The absolute magnitude $M$ of a supernova \cite{ref71,ref72} is obtained as :
\begin{center}
	\begin{equation}\label{eq35}
		M=16.08-25+5log_{10}(H_{0}/.026c)
	\end{equation}
	\par\end{center}
This provide the  apparent magnitude $m_b$ as

\begin{equation}\label{eq36}
	m_{b}=16.08+ 5log_{10}\left[\frac{1+z}{.026} \int^z_0\frac{dz}{h(z)}\right].
\end{equation}

\subsection{ $\chi^2$ for Apparant Magnitude $m_b$ and Distance Modulus `$\mu$':}
We use the most recent compilation of Supernovae pantheon samples which includes $715$ SN Ia plus $40$ SN Ia bined data in the range of( $ 0.01 \le  z \le 1.17$ ) \cite{ref73,ref74}. In $(mb, z)$ pairs, the pantheon compilation of SN Ia data is displayed. Here, we define $\chi^2$ for the parameters with the likelihood given by $ \varphi \propto e^{-\chi^2}$ in order to restrict the various parameters of the Universe in the resultant model.

$$ \chi^2= \frac{(mb_ {ob} - mb_{th} )^2}{mb_{err}^2}$$\\
and
$$ \chi^2= \frac{(\mu_ {ob} -\mu_{th} )^2}{\mu_{err}^2}$$ \\

The estimated present values of $H0$, $\Omega_m$, and parameter 'n' are displayed in the following table based on the minimum $\chi^2$.
\begin{table}[H]
	\centering
	\begin{tabular}{|c|c|}
		\hline
		Parameter          & Values    \\
		\hline
		$\Omega_{m0}$      &   0.40543 \\
		
		$\Omega_{\beta0}$ &  0.44068\\
		
		$n$                &     1.09094    \\
		
		$\chi^2$            & 23506.4  \\
		
		\hline
	\end{tabular}
	\caption{The best fit values of $\Omega_{m0}$  and `n' for the SN Ia data set of Apparent magnitude $m_b$ .}\label{T3}
\end{table}
\hspace{1mm}	
\begin{table}[H]
	\centering
	\begin{tabular}{|c|c|}
		\hline
		Parameter          & Values    \\
		\hline
		\\
		$H_{0}$   & 69.5047$_{-0.14054} ^{+0.14846} $\\
		\\
		$\Omega_{m0}$      &  0.291379$	_{-0.0448}^{+0.0478}$ \\
		\\
		$\Omega_{\beta0}$ & 0.678478\\
			$n$                &  1.01554  \\
		$\chi^2$            & 1220.59   \\
		\hline
	\end{tabular}
	\caption{The best fit values of $\Omega_{m0}$ and ' n` for the SN Ia data set of Distance modulus $\mu$ .}\label{T4}\end{table}

\begin{figure}[H]
	a.	\includegraphics[width=9cm,height=7cm,angle=0]{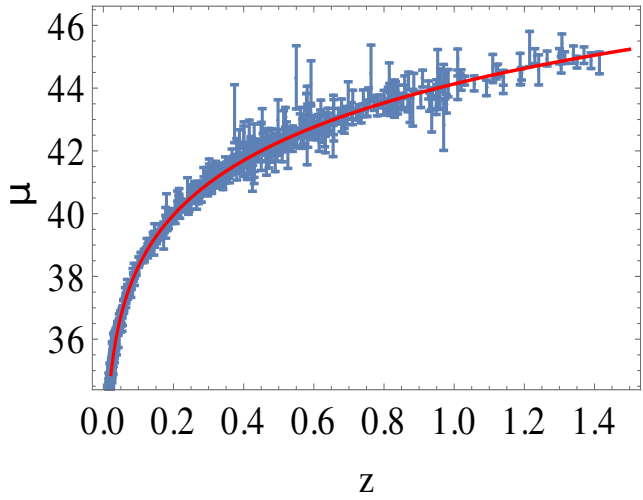}
	b.\includegraphics[width=9cm,height=7cm,angle=0]{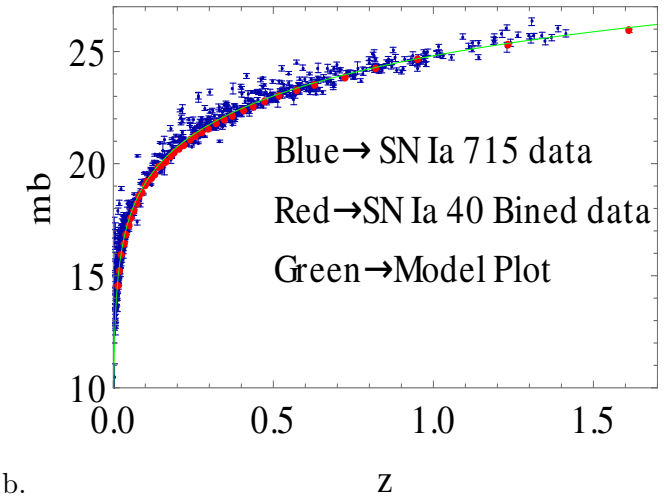}
	\caption{(a,b ) distance modulus ($\mu=M-m_b$) versus redshift ($z$) and Apparent magnitude ($m_b$) versus redshift ($z$) Error bar plots.}
\end{figure}
\begin{figure}[H]

a.\includegraphics[width=9cm,height=7cm,angle=0]{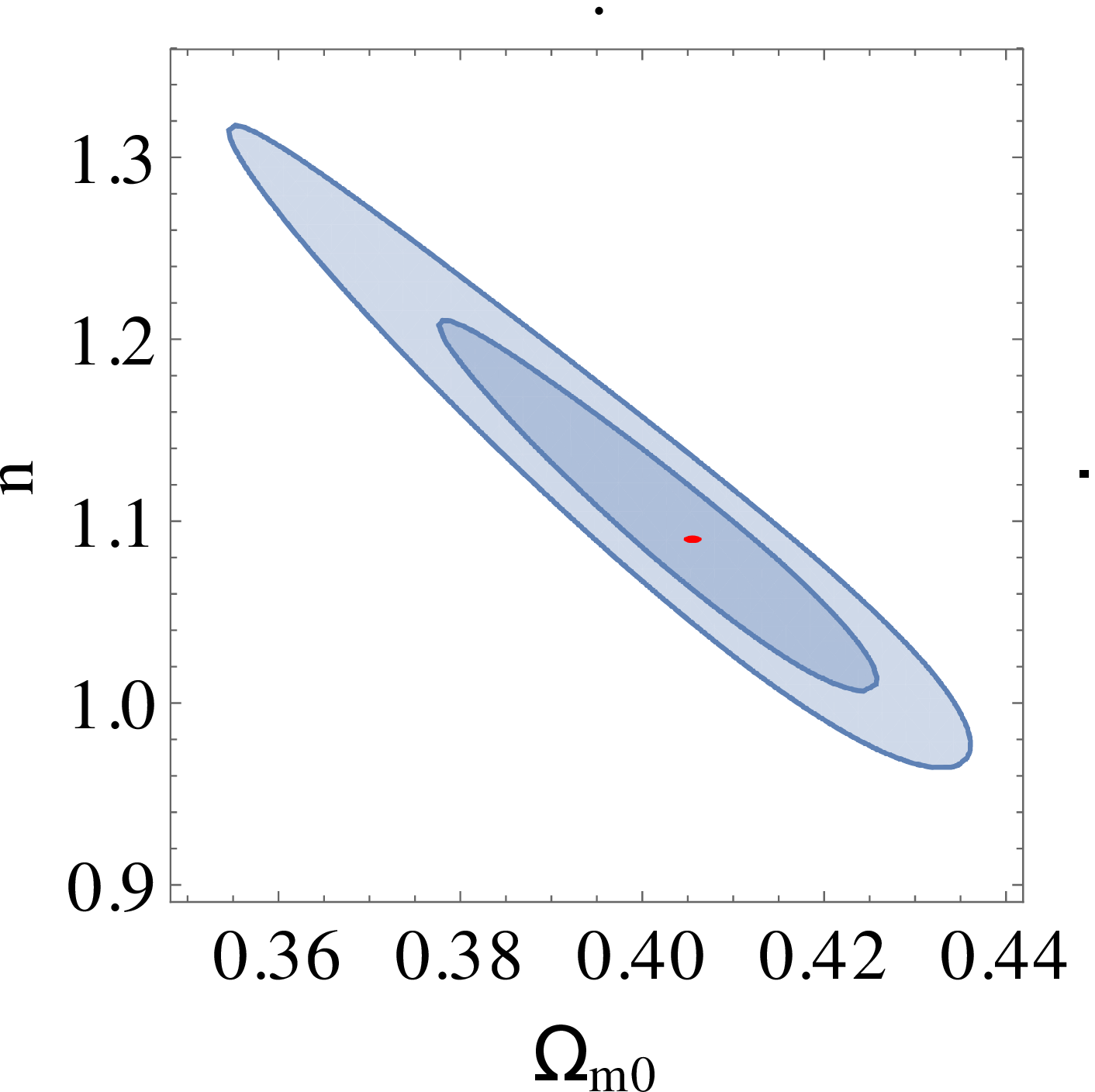}
b.\includegraphics[width=9cm,height=7cm,angle=0]{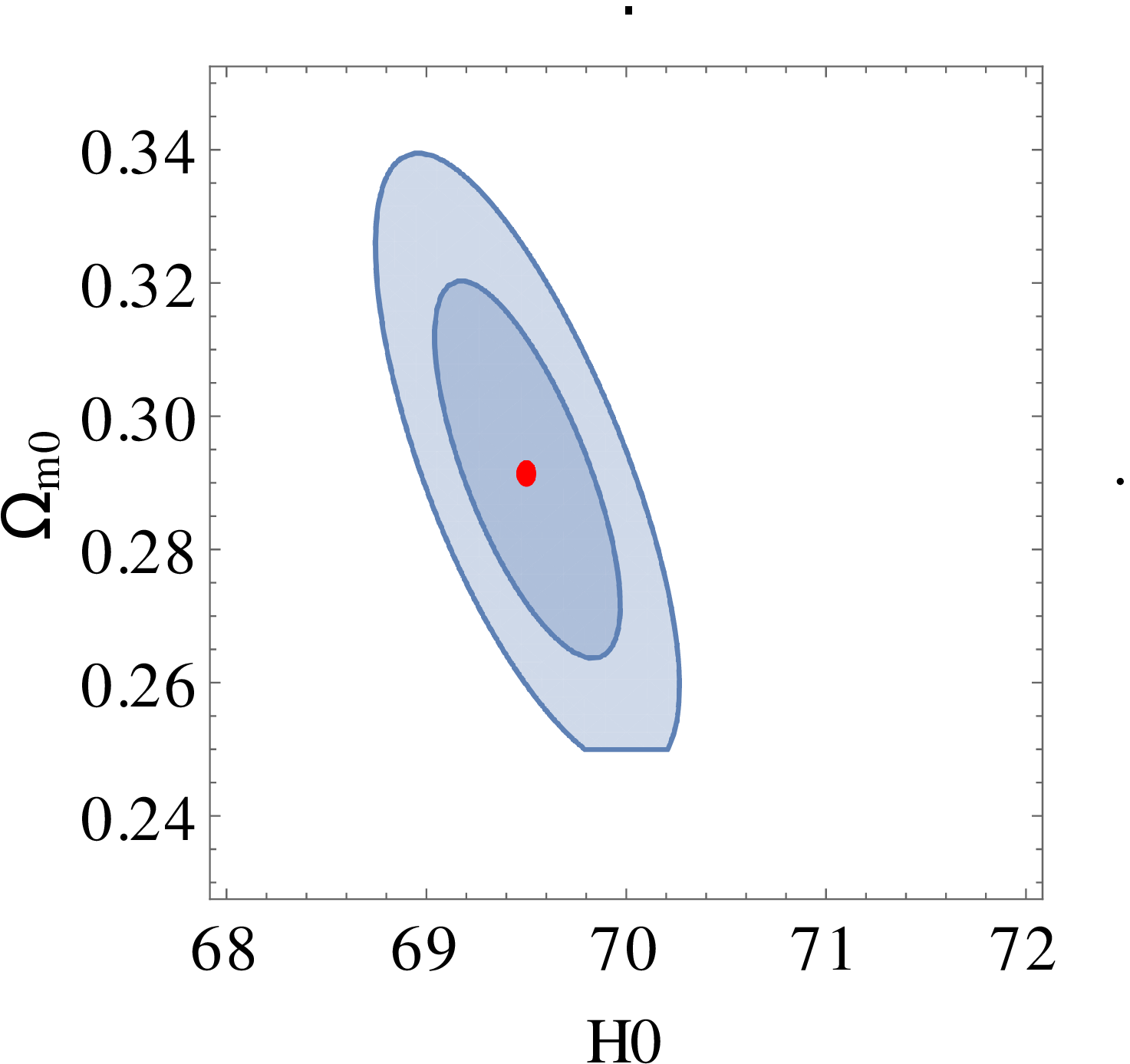}
c.\includegraphics[width=9cm,height=8cm,angle=0]{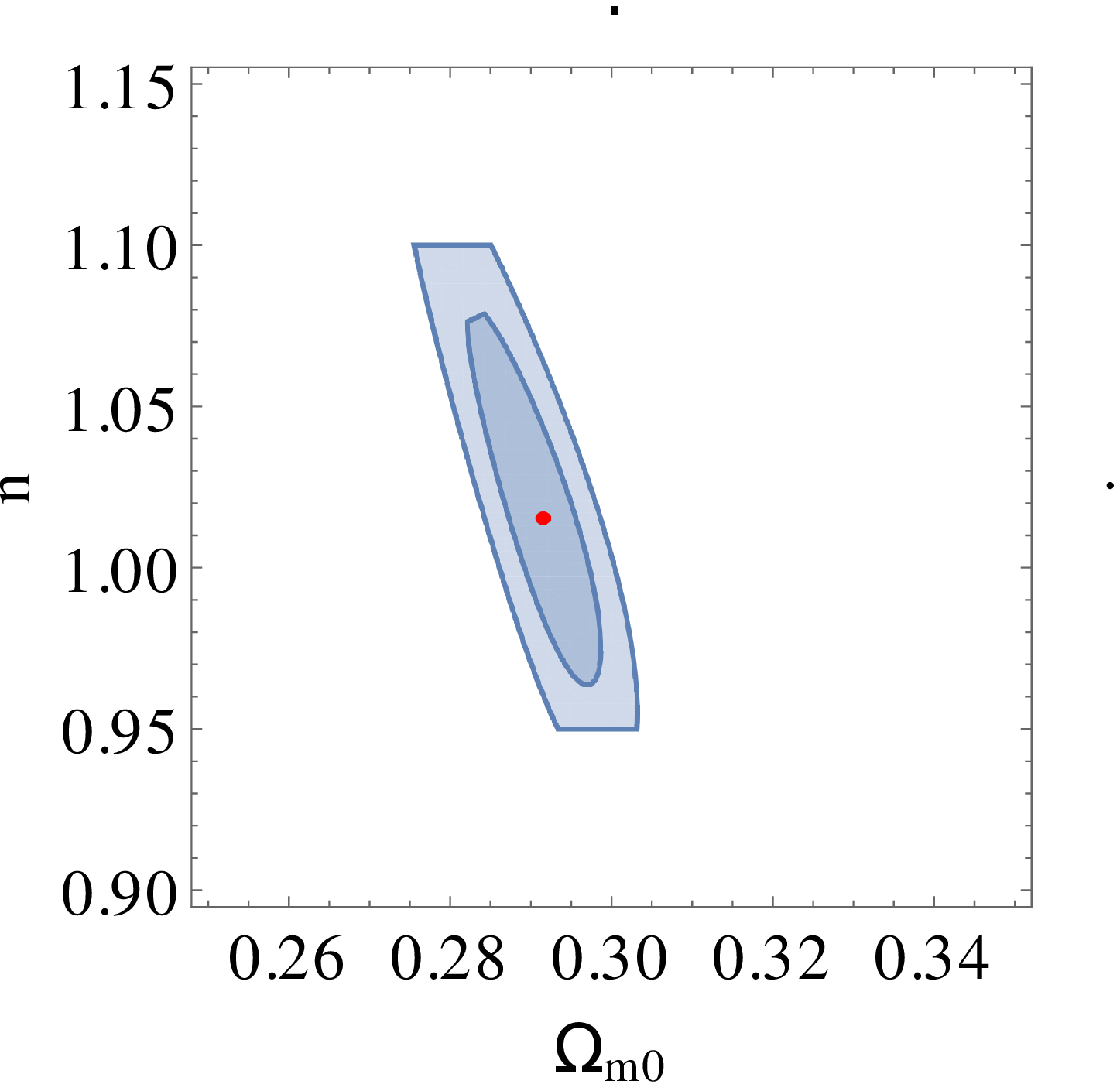}
d.\includegraphics[width=9cm,height=8cm,angle=0]{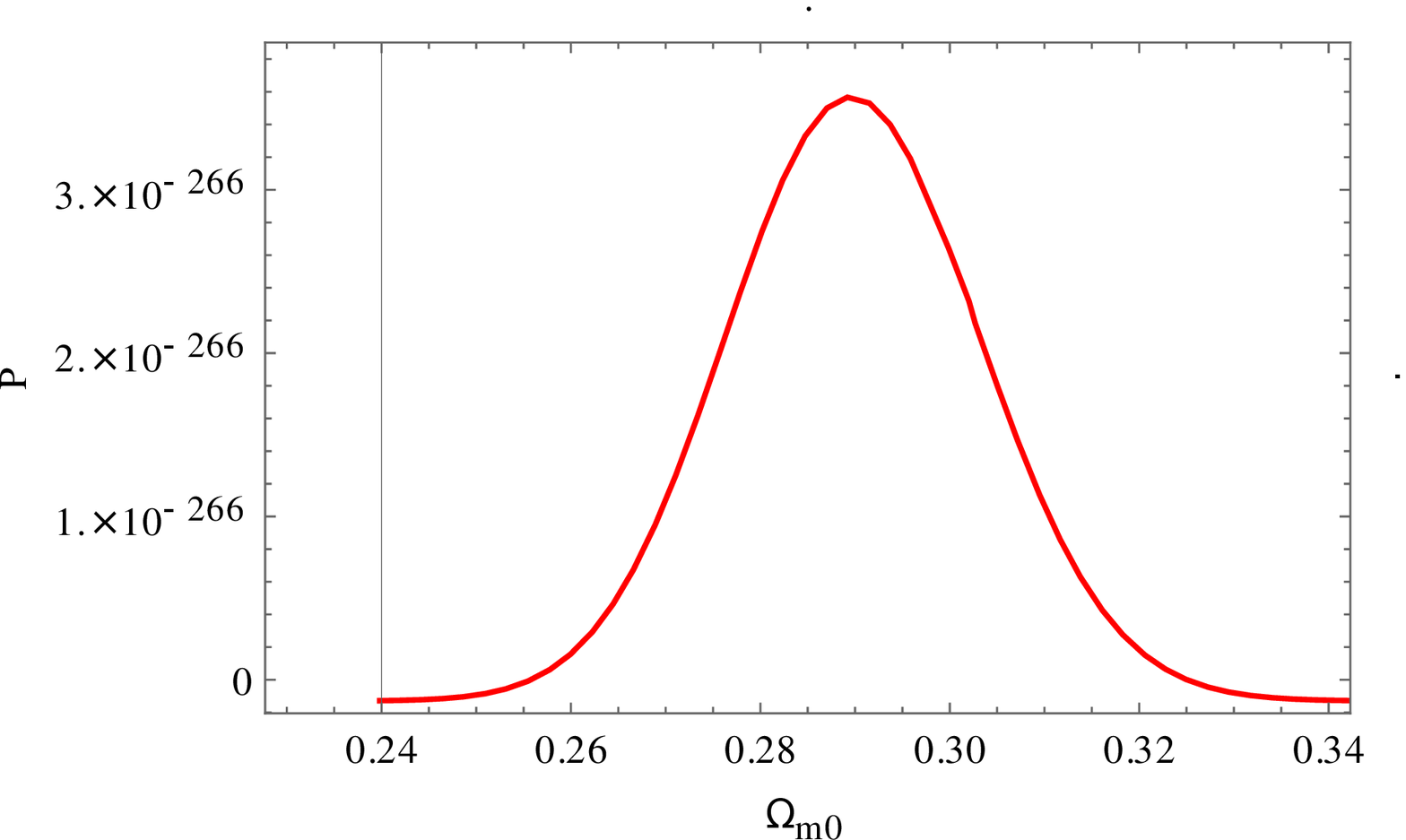}

	\caption{(a,b,c) $1\sigma$ and $2\sigma$ confidence regions for estimated parameters Hubble$H0$,  $\Omega_{m0}$  and `n' for Apparent magnitue and Distance modulus respectively.(d)Likelyhood plot barion energy parameter $\Omega_{m0}$ }
\end{figure}


\section*{Cosmographic Analysis}
Of all the rational approaches, cosmography has recently received a lot of attention \cite{ref76,ref77}. This model-independent method relies solely on the observational assumptions of the cosmological principle, allowing the study of dark energy evolution without the need to adopt a specific cosmological model \cite{ref78,ref79,ref80}. The usual space flight approach is based on Taylor's expansion of observations that can be directly compared to the data, and the results of such procedures are independent of the state equations assumed to study the evolution of the universe. For these reasons, cosmography has proven to be a powerful tool for breaking down the cosmological model, and is now widely used to understand the kinematics of the universe \cite{ref81}-\cite{ref113}.\\

The cosmological principle requires a scale factor as the only degree of freedom that governs the universe. Therefore, we can  expand the current Taylor series of $a(t)$ about present time and hence, we can define the cosmographic series coefficients like Hubble parameter $(H)$, deceleration parameter $(q)$, jerk $(j)$, snap $(s)$, lerk $(l)$ and max-out $(m)$  as given in \cite{ref75}:
\begin{equation}\label{eq37}
H=\frac{1}{a}\frac{da}{dt},~~q=-\frac{1}{aH^{2}}\frac{d^{2}a}{dt^{2}},~~j=\frac{1}{aH^{3}}\frac{d^{3}a}{dt^{3}}
\end{equation}
and
\begin{equation}\label{eq38}
s=\frac{1}{aH^{4}}\frac{d^{4}a}{dt^{4}},~~l=\frac{1}{aH^{5}}\frac{d^{5}a}{dt^{5}},~~m=\frac{1}{aH^{6}}\frac{d^{6}a}{dt^{6}}
\end{equation}
The dynamics of the late universe are studied using these quantities. The Hubble expansion's shape can be used to determine the physical characteristics of the coefficients. In particular, the sign of parameter $q$ tells us whether the cosmos is accelerating or decelerating. The sign of $j$ controls how the dynamics of the universe change, and positive values of $j$ denote the occurrence of transitional intervals when the universe's expansion changes. In order to distinguish between developing dark energy theories and cosmological constant behaviour, we also require the value of $s$. \\

Using the scale-factor (\ref{eq24})in (\ref{eq37}) \& (\ref{eq38}), we have found the cosmographic series coefficients $q, j, s, l, m$ as

\begin{equation}
 \label{eq39}
  q = -1+\frac{1}{k}~sech^{2}(k_{1}t)
\end{equation}

\begin{equation}\label{eq40}
j(t) = \frac{(k-1)(k-2)}{k^{2}}+\frac{3k-2}{k^{2}}~tanh^{2}(k_{1}t)
\end{equation}
\begin{equation}\label{eq41}
s(t) = \frac{(k-1)(k-2)(k-3)}{k^{3}}+\frac{2(k-1)(3k-4)}{k^{3}}tanh^{2}(k_{1}t)+\frac{3k-2}{k^{3}}tanh^{4}(k_{1}t)
\end{equation}

\begin{equation}\label{eq42}
l(t)  = \frac{(k-1)(k-2)(k-3)(k-4)}{k^{4}}+\frac{10(k-1)(k-2)^{2}}{k^{4}}tanh^{2}(k_{1}t)+\frac{15k^{2}-30k+16}{k^{4}}tanh^{4}(k_{1}t)
\end{equation}

\[
 m(t) = \frac{(k-1)(k-2)(k-3)(k-4)(k-5)}{k^{5}}+\frac{5(k-1)(k-2)(k-3)(3k-8)}{k^{5}}tanh^{2}(k_{1}t)
\]
\begin{equation}\label{eq43}
+\frac{(k-1)(45k^{2}-150k+136)}{k^{5}}tanh^{4}(k_{1}t)+\frac{15k^{2}-30k+16}{k^{5}}tanh^{6}(k_{1}t)
\end{equation}

where $k=\frac{2(2n-1)}{3n}$, $k_{1}=\frac{2}{3n}H_{0}\sqrt{\Omega_{\beta0}(2n-1)^{3}}$ and $t$ is the cosmic time.
\begin{table}[H]
  \centering
  \begin{tabular}{|c|c|c|c|}
     \hline
     Parameter & For $H_{0}$ data & Fo \bf{SN Ia 715 data  plus 40 bined data}& For \bf{SN Ia 715 data } \\
     \hline

      $H_{0}$ & 68.9596 & --       & 69.5047\\

     $q_{0}$ & -0.5211 & -0.3366   & -0.5561 \\

     $j_{0}$ & 0.9468 & 0.8469     & 0.9799 \\

     $s_{0}$ & -0.3106 & -0.6178   & -0.3157 \\

     $l_{0}$ & 2.854 & 3.208       & 2.896 \\

     $m_{0}$ & -9.504 & -13.25     & -9.594 \\

     $t_{0}$ & 13.5922 Gyrs & 12.1960 Gyrs & 13.8768\\

     $z_{t}$ & 0.6879  & 0.4553    & 0.7097 \\
     
     $t_{r}$ & 6.4912 Gyrs  & 9.78 Gyrs    & 5.8507 Gyrs \\
     
     \hline
   \end{tabular}
  \caption{The present values of cosmographic series coefficients $\{H_{0}, q_{0}, j_{0}, s_{0}, l_{0}, m_{0}\}$ for the best fit values of $\Omega_{m0}, \Omega_{\beta0}, H_{0}, n$ with three observational data sets as mentioned in Table 1, 3, 4.}\label{T5}
\end{table}
The equation (\ref{eq39}) shows the expression for the deceleration parameter (DP) $q(t)$ in terms of cosmic time $t$, and Figure 5a represents its geometrical behavior over redshift $z$. One can see that $q$ shows a signature-flipping (decelerating to accelerating phase) over its evolution with redshift at $z_{t}=0.6879, 0.4553, 0.7097$ along three data sets, i.e., the universe expansion is in decelerating phase for $z>z_{t}$ and it is in accelerating phase for $z<z_{t}$ which is closed to the recent observational values. The DP $q$ varies over $(-1, 0.5)$ and the present value of DP is estimated as $q_{0}=-0.5211, -0.3366, -0.5561$ (mentioned in Table $5$) which shows that our current universe is in accelerating phase and at a whole evolution of the model, it shows a transit phase universe model which is supported by the recent observations \cite{ref1}-\cite{ref10}. Figure 5b shows the geometrical behavior of best-fit Hubble function.\\

\begin{figure}[H]
	a.\includegraphics[width=9cm,height=8cm,angle=0]{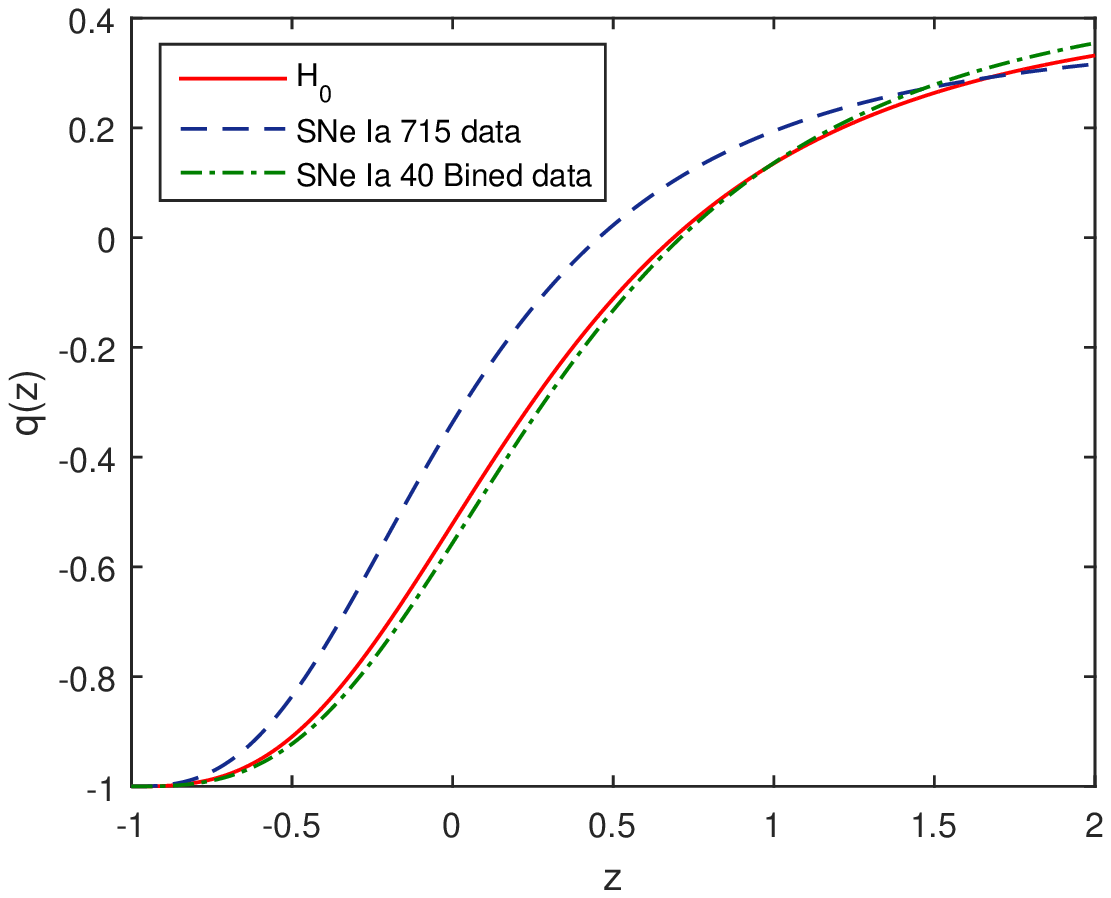}
    b.\includegraphics[width=9cm,height=8cm,angle=0]{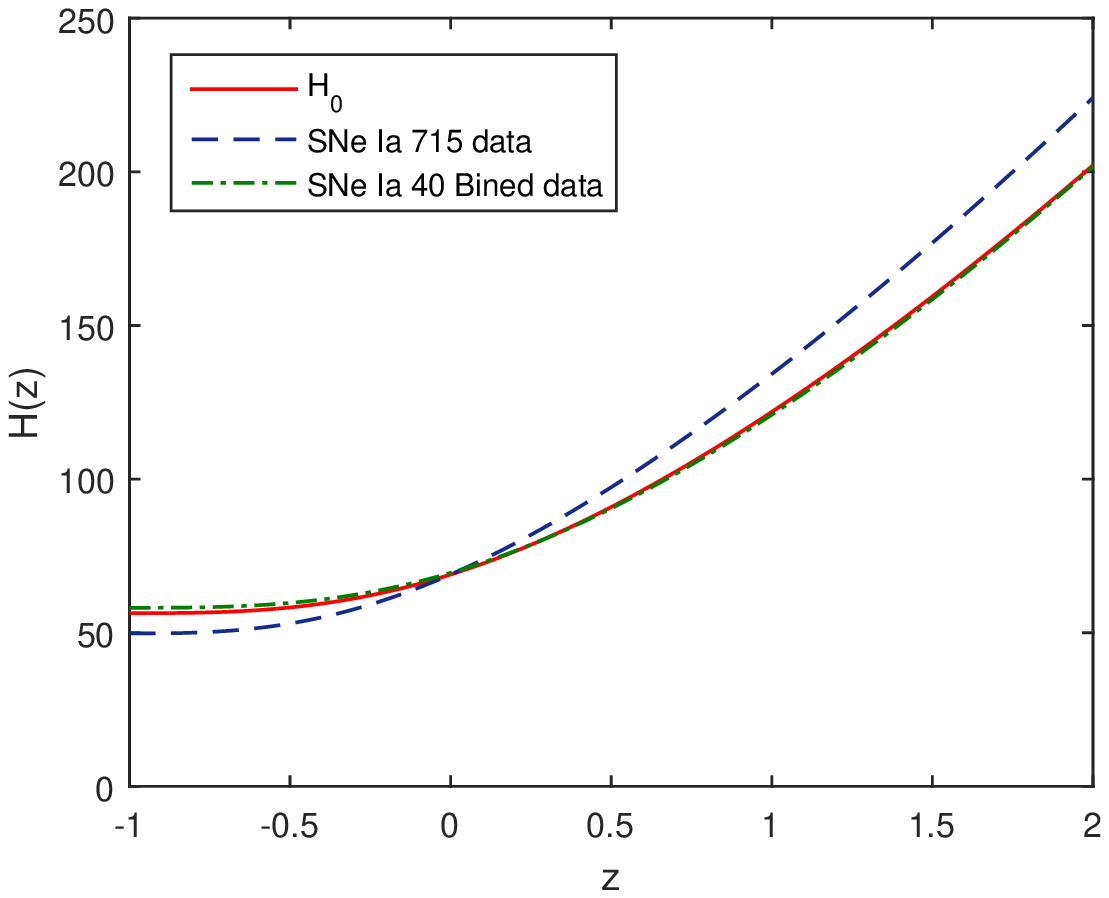}
   	\caption{(a) The plot of deceleration parameter $q(z)$ and (b) the plot of Hubble function $H(z)$ versus cosmic redshift $z$ for the best fit vales of $\Omega_{m0}, \Omega_{\beta0}, H_{0}, n$ along three data sets as mentioned in Table $1, 3, 4$.}
\end{figure}
\begin{figure}[H]
	a.\includegraphics[width=9cm,height=8cm,angle=0]{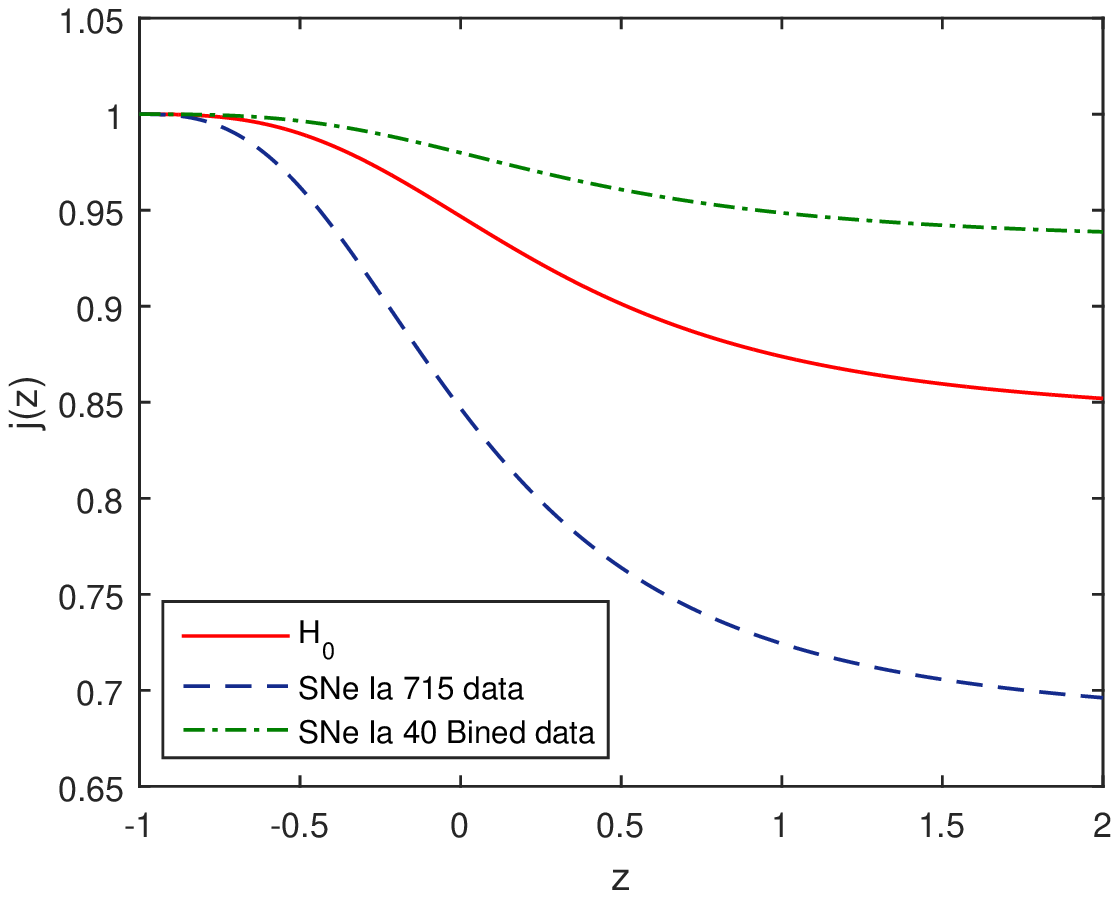}
    b.\includegraphics[width=9cm,height=8cm,angle=0]{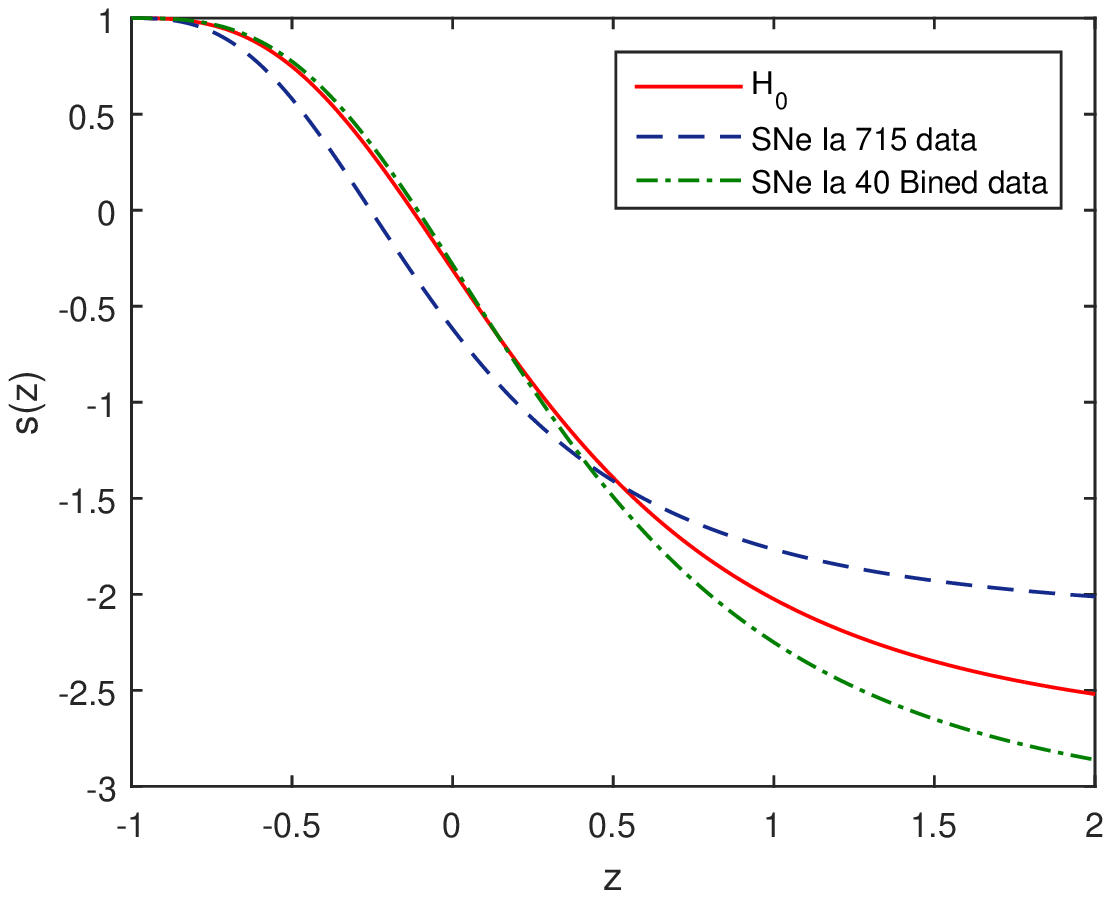}
    \caption{(a) The plot of jerk parameter $j(t)$ and (b)the plot of snap $s(z)$ versus cosmic redshift $z$ for the best fit vales of $\Omega_{m0}, \Omega_{\beta0}, H_{0}, n$ along three data sets as mentioned in Table $1, 3, 4$.}
\end{figure}
\begin{figure}[H]
	a.\includegraphics[width=9cm,height=8cm,angle=0]{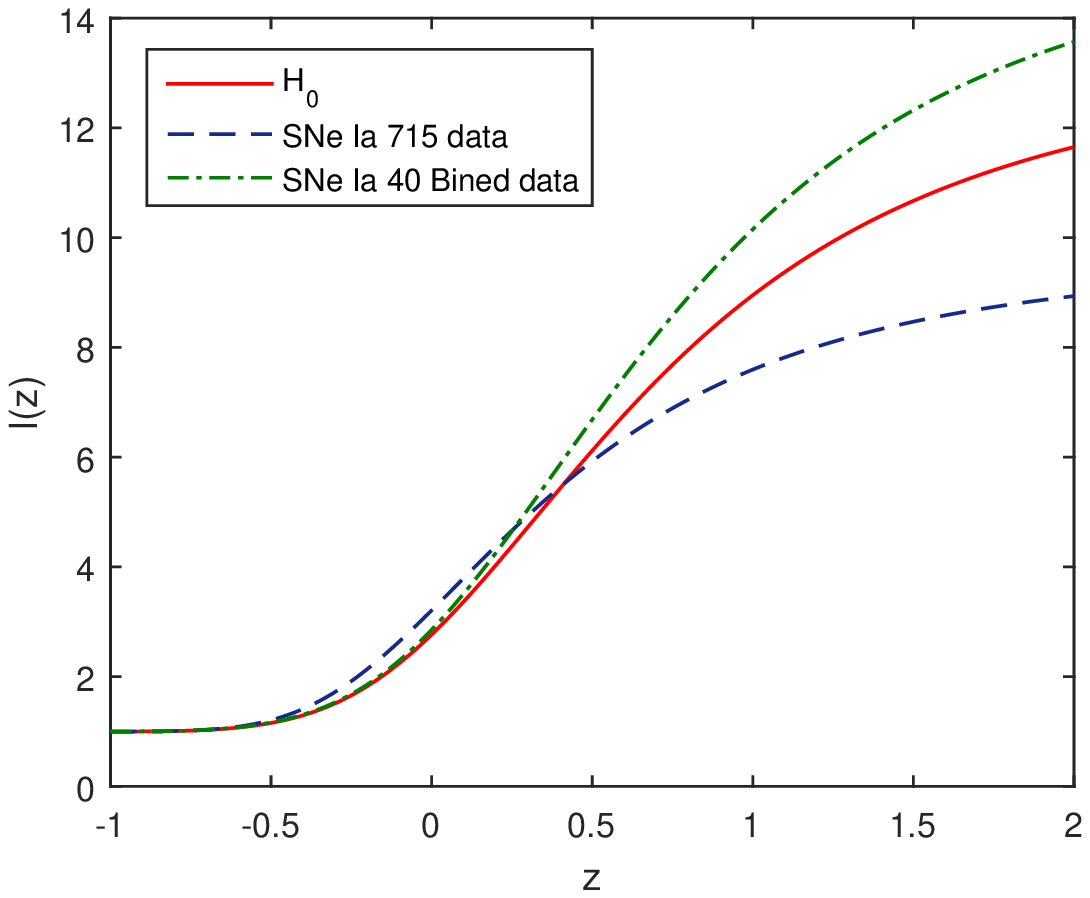}
    b.\includegraphics[width=9cm,height=8cm,angle=0]{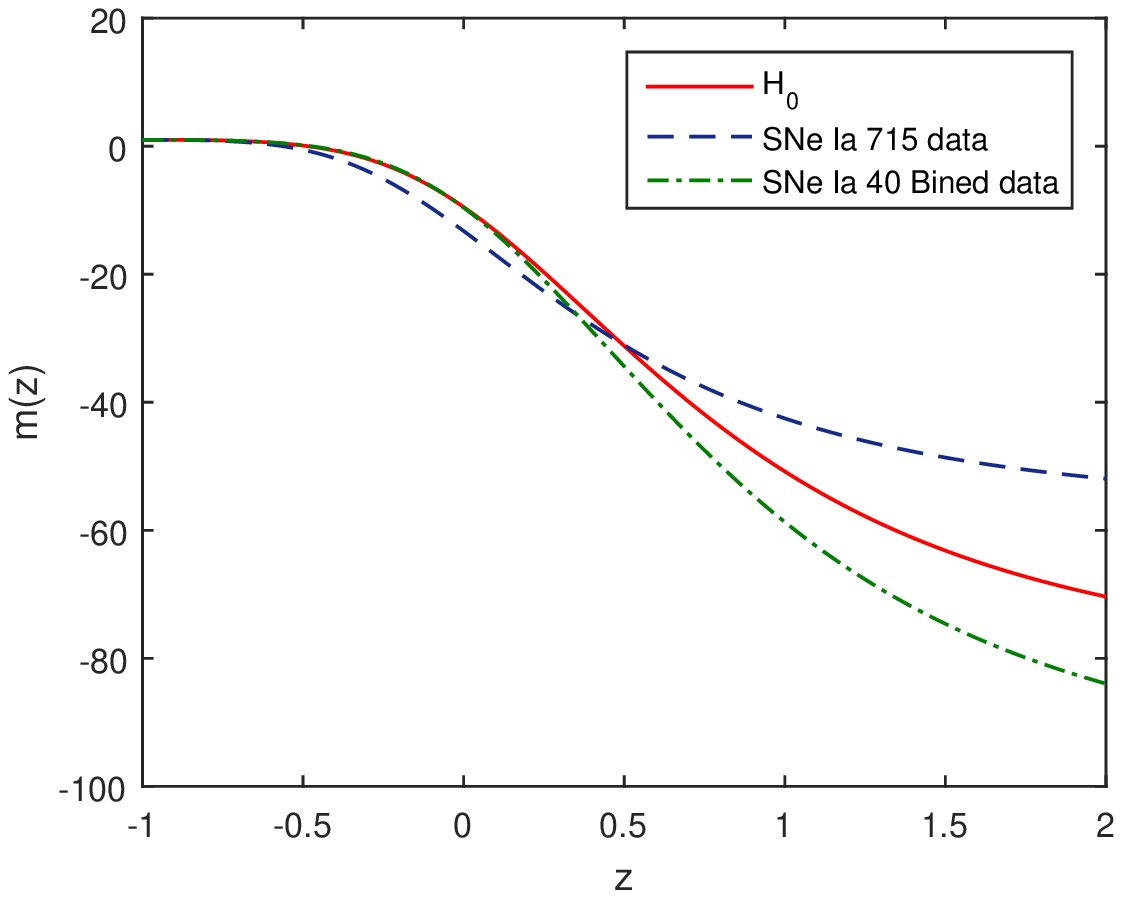}
	\caption{(a) The plot of lerk $l(z)$  and (b) the plot of max-out $m(z)$ versus cosmic redshift $z$ for the best fit vales of $\Omega_{m0}, \Omega_{\beta0}, H_{0}, n$ along three data sets as mentioned in Table $1, 3, 4$.}
\end{figure}
The next cosmographic coefficient is the jerk parameter $j(t)$ represented by the Eq.~(40), and its geometrical nature over redshift $z$ is shown in Figure $6a$. One can see that the value of the jerk parameter is always positive $(j>0)$, which indicates the occurrence of a transition time during which the universe modifies its expansion phase. The present value of jerk parameter in our model is estimated as $j_{0}=0.9468, 0.8469, 0.9799$ along three data sets and varies as $[0.852, 1]$ over redshift $-1\leq z\leq2$ (mentioned in Table $5$) \cite{ref75}. Thus, our derived model is a transit phase model supported by recent observations.\\

The next cosmographic coefficient is snap parameter $s$ represented by the Eq. (41), and its geometrical behavior is shown in Figure $6b$. The snap parameter shows the dark energy term or cosmological constant behavior of the model. The estimated present value of snap parameter is $s_{0}=-0.3106, -0.6178, -0.3157$ along three data sets which is supported by $\Lambda$CDM values. The other cosmographic coefficients are lerk $l(t)$, and max-out $m(t)$, whose expression is given by the Eqs.~(42) \& (43), respectively, and their geometrical behavior is represented by Figures $7a$ \& $7b$ respectively. The present values are estimated as $l_{0}=2.854, 3.208, 2.896$ and $m_{0}=-9.504, -13.25, -9.594$ along three data sets and varies with cosmic redshift over the range $[1, 11.65]$ \& $[-70.11, 1]$ respectively. Thus, one can see that as $t\to\infty$ (or $z\to-1$) then $\{q, j, s, l, m\}\to\{-1, 1, 1, 1, 1\}$ which is a good feature of our derived model.\\

The equation (\ref{eq20}) represents the expression for matter energy density in terms of scale factor and it may be shown in terms of redshift as $\rho=\rho_{0}(1+z)^{\frac{3}{2n-1}}$. We see that as $z\to0$ then $\rho\to\rho_{0}$ the present value.

\subsection*{Age of the present universe}
The age of the present universe can be calculated by
\begin{equation}\label{eq44}
t_{0}-t=-\int_{t_{0}}^{t}dt=\int_{0}^{z}\frac{dz}{(1+z)H(z)}
\end{equation}
Using (\ref{eq23}) in (\ref{eq44}) and integrating, we get
\begin{equation}\label{eq45}
H_{0}(t_{0}-t)=\frac{0.6923050296}{\sqrt{\Omega_{\beta0}(2n-1)}}~\left[ tanh^{-1}\left(\frac{\sqrt{\Omega_{m0}+\Omega_{\beta0}}}{\sqrt{\Omega_{\beta0}}}\right)-tanh^{-1}\left(\frac{\sqrt{\Omega_{m0}(1+z)^{2.8889}+\Omega_{\beta0}}}{\sqrt{\Omega_{\beta0}}}\right)\right]
\end{equation}
\begin{figure}[H]
	\centering
	\includegraphics[width=10cm,height=8cm,angle=0]{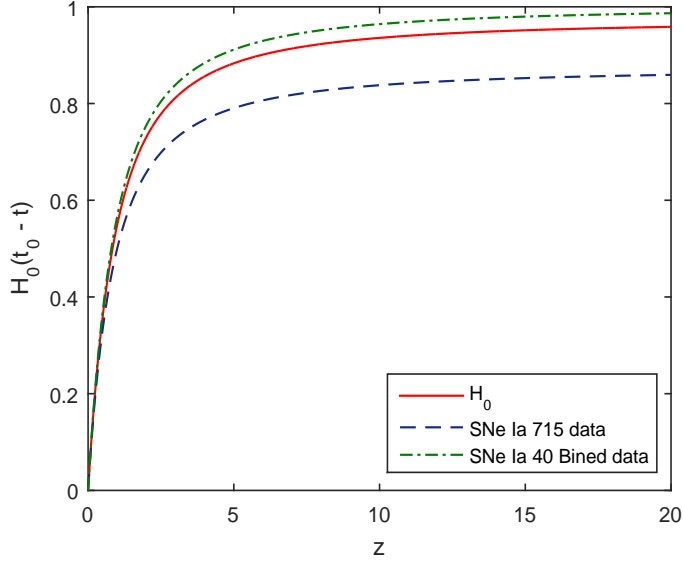}
   	\caption{The plot of cosmic time $H_{0}(t_{0}-t)$ versus redshift $z$ for the best fit vales of $\Omega_{m0}, \Omega_{\beta0}, H_{0}, n$ along three data sets as mentioned in Table $1, 3, 4$.}
\end{figure}
Figure $8$ shows the cosmic age of the universe over redshift $z$ and we have estimated the age of the present universe as $t_{0}=13.5922, 12.1960, 13.8768$ Gyrs respectively along three data sets which is very closed the the recent observational values in \cite{ref58,ref59,ref60}.
\section{Conclusions}
     This research paper deals with a transit dark energy cosmological model in $f(R, L_{m})$-gravity with observational constraints, where $R$ is Ricci scalar curvature and $L_{m}$ is matter Lagrangian for a perfect fluid. We have derived the field equations using a flat FLRW space-time metric and have found a relationship between energy density parameters $\Omega_{m0}, \Omega_{\beta0}$ through the Hubble function $H(z)$. We have approximated the present values of the energy parameters as $\Omega_{m0},~\Omega_{\beta0}$ and some other parameters by comparing observational and theoretical datasets using $R^{2}$ and $\chi^{2}$-test formula. The estimated present values of various parameters in the context of three data sets are as given in below Table $6$:
\begin{table}[H]
  \centering
  \begin{tabular}{|c|c|c|c|}
     \hline
     Parameter & For $H_{0}$ data & For\bf{ SN Ia 715 data}& For \bf{SN Ia 715 data  plus 40 bined data} \\
     \hline
      $\Omega_{m0}$ & 0.306973 & 0.40543 & 0.291379\\
      
      $\Omega_{\beta0}$ & 0.618952 & 0.44068 & 0.678478\\
      
     $n$              & 1.04     & 1.09094  & 1.01554\\
      
      $H_{0}$ & 68.9596 & 68.96       & 69.5047\\

     $q_{0}$ & -0.5211 & -0.3366   & -0.5561 \\

     $j_{0}$ & 0.9468 & 0.8469     & 0.9799 \\

     $s_{0}$ & -0.3106 & -0.6178   & -0.3157 \\

     $l_{0}$ & 2.854 & 3.208       & 2.896 \\

     $m_{0}$ & -9.504 & -13.25     & -9.594 \\

     $t_{0}$ & 13.5922 Gyrs & 12.1960 Gyrs & 13.8768\\

     $z_{t}$ & 0.6879  & 0.4553    & 0.7097 \\

     $t_{r}$ & 6.4912 Gyrs  & 9.78 Gyrs    & 5.8507 Gyrs \\

     \hline
   \end{tabular}
  \caption{The estimated present values of various parameters used in the above derived model over $0\leq z <3$.}\label{T6}
\end{table}

      The main features of the model are as follows:

      \begin{itemize}
        \item The derived model shows a signature-flipping point at $z=z_{t}$ (i.e. at this point the expansion phase of the universe is changed from the decelerating to the accelerating phase). Also, the model is decelerating for $z>z_{t}$ and it is in accelerating for $z<z_{t}$ (see Table $6$).
        \item The present value of deceleration parameter is measured as $q_{0}=-0.5211, -0.3366, -0.5561$ along three data sets which indicates that our present universe expansion is in accelerating phase supported by \cite{ref1}-\cite{ref10}.
        \item Also, we have measured the approximate present values of cosmographic series coefficients $\{H_{0}, q_{0}, j_{0}, s_{0}, l_{0}, m_{0}\}$ as mentioned in above Table $6$.
        \item Thus, we have found that as $t\to\infty$ (or $z\to-1$) then $\{q, j, s, l, m\}\to\{-1, 1, 1, 1, 1\}$ which indicates that our model shows $\Lambda$CDM model in late-time universe.
        \item The models shows a dark energy model and the $\beta$ behaves just like dark energy candidate and model parameter $n$ as scaling parameter.
        \item The cosmic age of the present universe is approximated as $t_{0}=13.5922, 12.1960, 13.8768$ Gyrs respectively along three data sets.
        \item Our universe has been come in accelerating phase $t_{r}$ Gyrs ago from today and the values of $t_{r}$ are mentioned in above Table $6$.
      \end{itemize}
 Thus our derived model shows good features of the recent cosmological models which will attract the researchers in this field for future investigations. 
\section*{Acknowledgement}
The Inter-University Centre for Astronomy and Astrophysics (IUCAA), Pune, India is gratefully acknowledged by the authors (A. Pradhan \& G. K. Gowami) for providing the resources and assistance necessary to finish this work.
The author (DCM) is thankful to the Center for Theoretical Physics and Mathematics, IASE (Deemed to be University), Churu, Rajasthan, India for providing facilities and support where part of this work is carried out.
	

\begin{thebibliography}{}
\bibitem {ref1}
A. G. Riess {\it et al.}, Observational evidence from supernovae for an accelerating universe
and a cosmological constant, \textit{Astron. J.} \textbf{116} (1998) 1009.
\bibitem {ref2}
S. Perlmutter {\it et al.}, Measurements of omega and lambda from 42 high-redshift supernovae,
\textit{Astrophys. J.} \textbf{517} (1999) 565.
\bibitem {ref3}
A. G. Riess et al., Type-Ia supernova discoveries of $Z \ge 1$ from the Hubble space
telescope: Evidence from past deceleration and constraints on dark energy evolution,
\textit{Astrophys. J.} \textbf{607} (2004) 665-687.
\bibitem {ref4}
D. J. Eisenstein, \textit{et al.}, Detection of the baryon acoustic peak in the large-scale correlation function of SDSS luminous red galaxies, \textit{Astrophys. J.} \textbf{633} (2005) 560.
\bibitem {ref5}
W. J. Percival, \textit{et al.}, Baryon acoustic oscillations in the Sloan Digital Sky Survey data release 7 galaxy sample, \textit{Mon. Not. R. Astron. Soc.} \textbf{401} (2010) 2148.
\bibitem {ref6}
D. N. Spergel et al., First-year Wilkinson Microwave Anisotropy Probe (WMAP)
observations:Determination of Cosmological parameters, \textit{Astrophys. J. Suppl. Ser.} \textbf{148}, (2003) 175-194. astro-ph/0302209.
\bibitem {ref7}
T. Koivisto, D. F. Mota, Dark energy anisotropic stress and large scale structure formation, \textit{Phys. Rev. D} \textbf{73} (2006) 083502.
\bibitem {ref8}
S. F. Daniel, Large scale structure as a probe of gravitational slip, \textit{Phys. Rev. D} \textbf{77} (2008) 103513.
\bibitem {ref9}
C. L. Bennett et al., First-year Wilkinson Microwave Anisotropy Probe (WMAP)
observations: Preliminary maps and basic results, \textit{Astrophys. J. Suppl. Ser.} \textbf{148}
(2003) 1-27.
\bibitem {ref10}
R. R. Caldwell, M. Doran, Cosmic microwave background and supernova constraints on quintessence: concordance regions and target models, \textit{Phys. Rev. D} \textbf{69} (2004) 103517.
\bibitem {ref11}
S. Weinberg, The cosmological constant problem, \textit{Rev. Mod. Phys.} \textbf{61} (1989) 1.
\bibitem {ref12}
E. J. Copeland, \textit{et al.}, Dynamics of dark energy, \textit{Int. J. Mod. Phys. D} \textbf{15} (2006) 1753.
\bibitem {ref13}
H. A. Buchdahl, Non-linear Lagrangians and cosmological theory, \textit{Mon. Not. R. Astron. Soc.} \textbf{150} (1970) 1.
\bibitem {ref14}
R. Kerner, Cosmology without singularity and nonlinear gravitational Lagrangians, \textit{Gen. Relativ. Gravit.} \textbf{14} (1982) 453.
\bibitem {ref15}
H. Kleinert, H. J. Schmidt, Cosmology with Curvature-Saturated Gravitational Lagrangian $R/\sqrt{1+l^{4}R^{2}}$, \textit{Gen. Relativ. Gravit.} \textbf{34} (2002) 1295.
\bibitem {ref16}
S. M. Carroll, \textit{et al.}, Is cosmic speed-up due to new gravitational physics?, \textit{Phys. Rev. D} \textbf{70} (2004) 043528.
\bibitem {ref17}
S. Capozziello, \textit{et al.}, Cosmological viability of $f(R)$-gravity as an ideal fluid and its compatibility with a matter dominated phase, \textit{Phys. Lett. B} \textbf{639} (2006) 135.
\bibitem {ref18}
L. Amendola, D. Polarski, S. Tsujikawa, Are $f(R)$ Dark Energy Models Cosmologically Viable?, \textit{Phys. Rev. Lett.} \textbf{98} (2007) 131302.
\bibitem {ref19}
S. Nojiri, S. D. Odintsov, Modified gravity with negative and positive powers of curvature: Unification of inflation and cosmic acceleration, \textit{Phys. Rev. D} \textbf{68} (2003) 123512.
\bibitem {ref20}
V. Faraoni, Solar system experiments do not yet veto modified gravity models, \textit{Phys. Rev. D} \textbf{74} (2006) 023529.
\bibitem {ref21}
P. J. Zhang, Behavior of $f(R)$ gravity in the solar system, galaxies, and clusters, \textit{Phys. Rev. D} \textbf{76} (2007) 024007.
\bibitem {ref22}
L. Amendola, S. Tsujikawa, Phantom crossing, equation-of-state singularities, and local gravity constraints in $f(R)$ models, \textit{Phys. Lett. B} \textbf{660} (2008) 125.
\bibitem {ref23}
S. Tsujikawa, Observational signatures of $f(R)$ dark energy models that satisfy cosmological and local gravity constraints, \textit{Phys. Rev. D} \textbf{77} (2008) 023507.
\bibitem {ref24}
T. Liua, X. Zhanga, W. Zhaoa, Constraining $f(R)$ gravity in solar system, cosmology and binary pulsar systems, \textit{ Phys. Lett. B} \textbf{777} (2018) 286-293.
\bibitem {ref25}
S. Capozziello, S. Tsujikawa, Solar system and equivalence principle constraints on $f(R)$ gravity by the chameleon approach, \textit{Phys. Rev. D} \textbf{77} (2008) 107501.
\bibitem {ref26}
S. M. Carroll, \textit{et al.}, Is cosmic speed-up due to new gravitational physics?, \textit{Phys. Rev. D} \textbf{70} (2004) 043528.
\bibitem {ref27}
A. A. Starobinsky, Disappearing cosmological constant in $f(R)$ gravity, \textit{JETP Lett.}\textbf{ 86} (2007) 157-163.
\bibitem {ref28}
S. Nojiri, S. D. Odintsov, Unifying inflation with $\Lambda$CDM epoch in modified $f(R)$ gravity consistent with Solar System tests, \textit{Phys. Lett. B} \textbf{657} (2007) 238.
\bibitem {ref29}
S. Nojiri, S. D. Odintsov, Modified $f(R)$ gravity unifying $R^{m}$ inflation with the $\Lambda$ epoch, \textit{Phys. Rev. D} \textbf{77} (2008) 026007.
\bibitem {ref30}
G. Cognola, \textit{et al.}, Class of viable modified $f(R)$ gravities describing inflation and the onset of accelerated expansion, \textit{Phys. Rev. D} \textbf{77} (2008) 046009.
\bibitem {ref31}
J. Santos, \textit{et al.}, Energy conditions in $f(R)$ gravity, \textit{Phys. Rev. D} \textbf{76} (2007) 083513.
\bibitem {ref32}
S. Capozziello, V. F. Cardone, V. Salzano, Cosmography of $f(R)$ gravity, \textit{Phys. Rev. D} \textbf{78} (2008) 063504.
\bibitem {ref33}
R. C. Nunes, \textit{et al.}, New observational constraints on $f(R)$ gravity from cosmic chronometers, \textit{J. Cosmol. Astropart. Phys.} \textbf{2017}(01) (2017) 005.
\bibitem {ref33a}
G. C. Samanta, N. Godani, Validation of energy conditions in wormhole geometry within viable f(R) gravity, 
\textit{Europ. Phys. J. C} \textbf{79} (2019) 1.
\bibitem {ref33b}
N. Godani, G. C. Samanta, Traversable wormholes in f(R) gravity with constant and variable redshift-functions, 
\textit{New Astronomy} \textbf{80} (2020) 101399.
\bibitem {ref33c}
N. Godani, G. C. Samanta, Charged traversable wormholes in f(R) gravity, 
\textit{Int. J. Geom. Methods Mod. Phys.} \textbf{18} (2021) 2150098.
\bibitem {ref33d}
N. Godani, Charged thin-shell wormholes in f(R) gravity, 
\textit{Int. J. Geom. Methods Mod. Phys.} \textbf{19} (2022) 2250035.
\bibitem {ref34}
O. Bertolami, \textit{et al.}, Extra force in $f(R)$ modified theories of gravity, \textit{Phys. Rev. D} \textbf{75} (2007) 104016.
\bibitem {ref35}
T. Harko, Modified gravity with arbitrary coupling between matter and geometry, \textit{Phys. Lett. B} \textbf{669} (2008) 376.
\bibitem {ref36}
T. Harko, Galactic rotation curves in modified gravity with nonminimal coupling between matter and geometry, \textit{Phys. Rev. D} \textbf{81} (2010) 084050.
\bibitem {ref37}
T. Harko, The matter Lagrangian and the energy-momentum tensor in modified gravity with nonminimal coupling between matter and geometry, \textit{Phys. Rev. D} \textbf{81} (2010) 044021.
\bibitem {ref38}
S. Nesseris, Matter density perturbations in modified gravity models with arbitrary coupling between matter and geometry, \textit{Phys. Rev. D} \textbf{79} (2009) 044015.
\bibitem {ref39}
V. Faraoni, Viability criterion for modified gravity with an extra force, \textit{Phys. Rev. D} \textbf{76} (2007) 127501.
\bibitem {ref40}
V. Faraoni, Lagrangian description of perfect fluids and modified gravity with an extra force, \textit{Phys. Rev. D} \textbf{80} (2009) 124040.
\bibitem {ref41}
T. Harko, F. S. N. Lobo, $f(R, L_{m})$ gravity, \textit{Eur. Phys. J. C} \textbf{70} (2010) 373-379.
\bibitem {ref42}
V. Faraoni, Cosmology in Scalar-Tensor Gravity, Kluwer Academic, Dordrecht, 2004.
\bibitem {ref43}
O. Bertolami, J. Piramos, S. Turyshev, General Theory of Relativity: Will it survive the next decade?, {\it Lasers, Clocks and Drag-Free Control}, Springer, Berlin, Heidelberg, (2008) 27-74.
\bibitem {ref44}
J. Wang, K. Liao, Energy conditions in $f(R, L_{m})$ gravity, \textit{Class. Quantum Gravity} \textbf{29} (2012) 215016.
\bibitem {ref45}
B. S. Goncalves, P. H. R. S. Moraes, Cosmology from non-minimal geometry-matter coupling, arXiv:2101.05918.
\bibitem{ref46}
L. V. Jaybhaye {\it et al.}, Cosmology in $f(R, L_{m})$ gravity, \textit{Phys. Lett. B} {\bf 831} (2022) 137148.
\bibitem {ref47}
B. Ryden, Introduction to Cosmology, Addison Wesley, San Francisco, United States of America, 2003.
\bibitem {ref48}
T. Harko, F. S. N. Lobo, Generalized curvature-matter couplings in modified gravity, \textit{Galaxies} \textbf{2014}(2) (2014) 410-465.
\bibitem {ref49}
T. Harko, \textit{et al.}, Gravitational induced particle production through a nonminimal curvature-matter coupling, \textit{Eur. Phys. J. C} \textbf{75} (2015) 386.
\bibitem{ref50}
E. J. Copeland, M. Sami and S. Tsujikawa, Dynamic of dark energy, {\it Int. J. Mod. Phys. D} \textbf{15} (2006) 1753-1936.
\bibitem {ref51}
S. Agarwal, R.K. Pandey, A.Pradhan, LRS Bianchi type II perfect fluid cosmological models in normal gauge for Lyra’s manifold, {\it Int. J. Theor. Phys.} {\bf 50} (2011) 296.
\bibitem {ref52}
A. Pradhan, S. Agarwal, G.P. Singh, LRS Bianchi type-I universe in Barber’s second self creation theory, {\it Int. J. Theor. Phys.} {\bf 48} (2009) 158.
 \bibitem {ref53}
E. Macaulay, et al., First cosmological results using Type Ia supernovae from the dark energy survey: measurement of the Hubble constant, {\it Mon. Not. R. Astro. Soc.} {\bf 486} (2019) 2184.
\bibitem {ref54}
C. Zhang, et al., Four new observational H(z) data from luminous red galaxies in the sloan digital sky survey data release seven, {\it Res. Astron. Astrophys.} {\bf 14} (2014) 1221.
	
\bibitem {ref55}
D. Stern, et al., Cosmic chronometers: constraining the equation of state of dark energy.I: H(z) measurements, {\it J. Cosmol. Astropart. Phys.} {\bf 1002} (2010) 008.
	 	
\bibitem {ref56}
E. G. Naga, et al., Clustering of luminous red galaxiesIV. Baryon acoustic peak in the line-of-sight direction and a direct measurement of H(z), Mon. Not. R. Astro. Soc. {\bf 399} (2009) 1663.
\bibitem{ref57}	
R. Aurich and F. Steiner, Dark energy in a hyperbolic universe, \textit{Monthly Notices of the Royal Astronomical Society} \textbf{334} (2002) 735.
\bibitem{ref58}
Planck Collaboration, N. Aghanim, Y. Akrami, et al. Planck 2018 results. VI. Cosmological
parameters, {\it A \& A} {\bf 641} (2020) A6.
\bibitem{ref59}
A. G. Riess, S. Casertano, W. Yuan, et al., (2021) Cosmic distances calibrated to $1\%$ precision with Gaia
EDR3 parallaxes and Hubble Space Telescope photometry of 75 Milky Way Cepheids Confirm Tension with KCDM, {\it ApJ} {\bf 908} (2021) L6.
\bibitem{ref60}
J. V. Cunha, Kinematic constraints to the transition redshift from supernovae type
Ia union data, {\it Phys. Rev. D} \textbf{79} (2009) 047301.
  \bibitem {ref61}
	 D. H Chauang, Y. Wang, Modelling the anisotropic two-point galaxy correlation function on small scales and single-probe measurements of H(z), DA(z) and f(z) $\sigma$8(z) from the sloan digital sky survey DR7 luminous red galaxies,
	 {\it Mon. Not. R. Astro. Soc.} {\bf 435} (2013) 255.
	
	 \bibitem {ref62}
	 S. Alam, et al., The clustering of galaxies in the completed SDSS -III Baryon Oscillation Spectroscopic Survey: cosmological
	 analysis of the DR12 galaxy sample, {\it Mon. Not. R. Astron. Soc.} {\bf 470} (2017) 2617.
	 	
	 \bibitem {ref63}
	 A. L. Ratsimbazafy, et al., Age-dating luminous red galaxies observed with the Southern African Large Telescope, {\it Mon. Not. R. Astron. Soc.} {\bf 467} (2017) 3239.
	 \bibitem {ref64}
	 L. Anderson, et al., The clustering of galaxies in the SDSS-III Baryon oscillation Spectro-scopic Survey: baryon acoustic oscillations in the data releases 10 and 11 galaxy samples, {\it Mon. Not. R. Astron. Soc.} {\bf 441} (2014) 24.
	 \bibitem {ref65}
	 M. Moresco, Raising the bar: new constraints on the Hubble parameter with cosmic chronometers at z $\equiv$ 2, {\it Mon. Not. R. Astron. Soc.} {\bf 450} (2015) L16.
	\bibitem {ref66}
	N. G. Busa, et al., Baryon acoustic oscillations in the Ly$\alpha$ forest of BOSS quasars, {\it Astron. $\&$ Astrophys.} {\bf 552} (2013) A96.

	\bibitem {ref67}
	M. Moresco, et al., Improved constraints on the expansion rate of the Universe up to $z\sim 1.1$ from the spectroscopic evolution of
	cosmic chronometers, {\it J. Cosmol. Astropart. Phys.} {\bf 2012} (2012) 006.
		
	\bibitem {ref68}
	J. Simon, L. Verde, R. Jimenez, Constraints on the redshift dependence of the dark energy potential, {\it Phys. Rev. D} {\bf 71}
	(2005) 123001.

	\bibitem {ref69}
	M. Moresco et al., A 6 $\%$ measurement of the Hubble parameter at z $\sim 0.45$ direct evidence of the epoch of cosmic re-acceleration,
	{\it J. Cosmol. Astropart. Phys.}  {\bf 05} (2016) 014.

	\bibitem {ref70}
	 G.F.R. Ellis, M.A.H. MacCallum, A class of homogeneous cosmological models, {\it Commun. Math. Phys.} {\bf 12} (1969) 108.

           \bibitem{ref71}
           G. K. Goswami, R. N. Dewangan, A. K. Yadav, {\it Astrophys. Space Sci.} {\bf 361} (2016) 119.

           \bibitem{ref72}
            G. K. Goswami, R. N. Dewangan, A. K. Yadav, A. Pradhan, {\it Astrophys. Space Sci.} {\bf 361} (2016) 47.

         \bibitem {ref73}
         A. K. Camlibel, I. Semiz, M. A. Feyizoglu, Pantheon update on a model-independent analysis of cosmological supernova data, {\it Class. Quantum Grav.} 37 (2020) 235001.

         \bibitem {ref74}
         D. M. Scolnic {\it et al.}, The complete light-curve sample of spectroscopically confirmed SNe Ia from Pan$-$STARRS1 and cosmological constraints from the combined pantheon sample, {\it Astrophys. J.} 859 (2018) 101.

\bibitem {ref75}
S. Capozziello \textit{et al.}, Extended Gravity Cosmography, {\it Int. J. Mod. Phys. D} {\bf 28} (2019) 1930016. .
\bibitem{ref76}
M. Visser, Cosmography: Cosmology without the Einstein equations, \textit{Gen. Rel. Grav.} \textbf{37} (2005) 1541; M. Visser, Conformally Friedmann-Lemaitre-Robertson-Walker cosmologies, \textit{Class. Quant. Grav.} \textbf{32} (2015) 135007.
\bibitem{ref77}
P. K. S. Dunsby, O. Luongo, On the theory and applications of modern cosmography, \textit{Int. J. Geom. Meth. Mod. Phys.} \textbf{13} (2016) 1630002.
\bibitem{ref78}
E. R. Harrison, Observational tests in cosmology, \textit{Nature} \textbf{260} (1976) 591.
\bibitem{ref79}
O. Luongo, Dark energy from a positive jerk parameter, \textit{Phys. Lett. A} \textbf{28} (2013) 1350080.
\bibitem{ref80}
S. Capozziello, M. De Laurentis, O. Luongo, A. C. Ruggeri, Cosmographic constraints and cosmic fluids, \textit{Galaxies} \textbf{1} (2013) 216.
\bibitem{ref81}
A. Mukherjee, N. Paul, H. K. Jassal, Constraining the dark energy statefinder hierarchy in a kinematic approach, \textit{J. Cosm. Astrop. Phys.} \textbf{1901} (2019) 005.
\bibitem{ref82}
S. Capozziello, O. Farooq, O. Luongo, B. Ratra, Cosmographic bounds on the cosmological deceleration-acceleration transition redshift in $f(R)$ gravity, \textit{Phys. Rev D} \textbf{90} (2014) 044016.
\bibitem{ref83}
A. Aviles, A. Bravetti, S. Capozziello, O. Luongo, Updated constraints on $f(R)$ gravity from cosmography, \textit{Phys. Rev. D} \textbf{87} (2013) 044012.
\bibitem{ref84}
S. Capozziello, M. De Laurentis, O. Luongo, Connecting early and late universe by $f(R)$ gravity, \textit{Int. J. Mod. Phys. D} \textbf{24} (2015) 1541002.
\bibitem{ref85}
S. Capozziello, O. Luongo, Information entropy and dark energy evolution, \textit{Int. J. Mod. Phys. D} \textbf{27} (2018) 1850029.
\bibitem{ref86}
O. Luongo, H. Quevedo, Cosmographic study of the universe's specific heat: a landscape for Cosmology?, \textit{Gen. Rel. Grav.} \textbf{46} (2014) 1649.
\bibitem{ref87}
H. Velten, S. Gomes, V. C. Busti, Gauging the cosmic acceleration with recent type Ia supernovae data sets, \textit{Phys. Rev. D} \textbf{97} (2018) 083516.
\bibitem{ref88}
U. Andrade, C. A. P. Bengaly, J. S. Alcaniz, B. Santos, Isotropy of low redshift type Ia supernovae: A Bayesian analysis, \textit{Phys. Rev. D} \textbf{97} (2018) 083518.
\bibitem{ref89}
C. Rodrigues Filho and Edesio M. Barboza, Constraints on kinematic parameters at $z\neq0$, \textit{J. Cosm. Astrop. Phys.} \textbf{1807} (2018) 037.
\bibitem{ref90}
O. Luongo, G. B. Pisani, A. Troisi, Cosmological degeneracy versus cosmography: A cosmographic dark energy model, \textit{Int. J. Mod. Phys. D} \textbf{26} (2017) 1750015.
\bibitem{ref91}
 Z. Y. Yin, H. Wei, Observational constraints on growth index with cosmography, \textit{Eur. Phys. J. C} \textbf{79} (2019) 698.
\bibitem{ref92}
A. Al Mamon, K. Bamba, Observational constraints on the jerk parameter with the data of the Hubble parameter, \textit{Eur. Phys. J. C} \textbf{78} (2018) 862.
\bibitem{ref93}
A. Piloyan, S. Pavluchenko, L. Amendola, Limits on the Reconstruction of a Single Dark Energy Scalar Field Potential from SNe Ia Data, \textit{Particles} \textbf{1} (2018) 23.
\bibitem{ref94}
F. Montanari, S. Rasanen, Backreaction and FRW consistency conditions, \textit{J. Cosm. Astrop. Phys.} \textbf{1711} (2017) 032.
\bibitem{ref95}
X. B. Zou, H. K. Deng, Z. Y Yin, H. Wei, Model-independent constraints on Lorentz invariance violation via the cosmographic approach, \textit{Phys. Lett. B} \textbf{776} (2018) 284.
\bibitem{ref96}
W. Yang, L. Xu, H. Li, Y. Wu, J. Lu, Testing the Interacting Dark Energy Model with Cosmic Microwave Background Anisotropy and Observational Hubble Data, \textit{Entropy} \textbf{19} (2017) 327.
\bibitem{ref97}
C. S. Carvalho, S. Basilakos, Angular distribution of cosmological parameters as a probe of inhomogeneities: a kinematic parametrisation, \textit{Astron. Astrophys.} \textbf{592} (2016) A152.
\bibitem{ref98}
O. Luongo, H. Quevedo, Self-accelerated Universe Induced by Repulsive Effects as an Alternative to Dark Energy and Modified Gravities, \textit{Found. Phys.} \textbf{48} (2018) 17-26.
\bibitem{ref99}
I. Semiz, A. K. Camlibel, What do the cosmological supernova data really tell us? \textit{J. Cosm. Astrop. Phys.} \textbf{1512} (2015) 038.
\bibitem{ref100}
Y. L. Bolotin, V. A. Cherkaski, O. A. Lemets, New cosmographic constraints on the dark energy and dark matter coupling, \textit{Int. J. Mod. Phys. D} \textbf{25} (2016) 1650056.
\bibitem{ref101}
B. Bochner, D. Pappas, M. Dong, Testing lambda and the limits of cosmography with the Union 2.1 Supernova Compilation, \textit{Astrophys. J.} \textbf{814} (2015) 7.
\bibitem{ref102}
S. Nesseris, J. Garcia-Bellido, Comparative analysis of model-independent methods for exploring the nature of dark energy, \textit{Phys. Rev. D} \textbf{88} (2013) 063521.
\bibitem{ref103}
O. Farooq, S. Crandall, B. Ratra, Binned Hubble parameter measurements and the cosmological deceleration-acceleration transition, \textit{Phys. Lett. B} \textbf{726} (2013) 72.
\bibitem{ref104}
F. A. Teppa Pannia, S. E. Perez Bergliaffa, Evolution of Vacuum Bubbles Embeded in Inhomogeneous Spacetimes, \textit{J. Cosm. Astrop. Phys.} \textbf{1308} (2013) 030.
\bibitem{ref105}
C. J. A. P. Martins, M. Martinelli, E. Calabrese, M. P. L. P. Ramos, Real-time cosmography with redshift derivatives, \textit{Phys. Rev. D} \textbf{94} (2016) 043001.
\bibitem{ref106}
F. Piazza, T. Schucker, Minimal cosmography, \textit{Gen. Rel. Grav.} \textbf{48} (2016) 41.
\bibitem{ref107}
K. Bamba, S. Capozziello, S. Nojiri, S. D. Odintsov, Dark energy cosmology: the equivalent description via different theoretical models and cosmography tests, \textit{Astrop. Sp. Sci.} \textbf{342} (2012) 155.
\bibitem{ref108}
S. Capozziello, V. F. Cardone, H. Farajollahi, A. Ravanpak, Cosmography in $f(T)$ gravity, \textit{Phys. Rev. D} \textbf{84} (2011) 043527.
\bibitem{ref109}
J. C. Carvalho, J. S. Alcaniz, Cosmography and cosmic acceleration, \textit{Mon. Not. Roy. Astron. Soc.} \textbf{418} (2011) 1873.
\bibitem{ref110}
M. Bouhmadi-Lopez, S. Capozziello, V. F. Cardone, Cosmography of $f(R)$-brane cosmology, \textit{Phys. Rev. D} \textbf{82} (2010) 103526.
\bibitem{ref111}
S. Capozziello, V. F. Cardone, V. Salzano, Cosmography of $f(R)$ gravity, \textit{Phys. Rev. D} \textbf{78} (2008) 063504.
\bibitem{ref112}
M. V. John, Cosmography, decelerating past, and cosmological models: learning the Bayesian way, \textit{Astrophys. J.} \textbf{630} (2005) 667.
\bibitem{ref113}
S. Capozziello, Ruchika, A. A. Sen, Model-independent constraints on dark energy evolution from low-redshift observations, \textit{Mon. Not. Roy. Astron. Soc.} \textbf{484} (2019) 4484.

\end{thebibliography}
\end{document}